\documentclass[superscriptaddress,groupedaddress,nofootnoteinbib,12pt]{article}  

\usepackage{array}
\usepackage{graphicx}  
\usepackage{dcolumn}   
\usepackage{bm}        
\usepackage{amssymb}   
\usepackage{yfonts}    
\usepackage{slashed}	 
\usepackage{tikz}			 
\usetikzlibrary{shapes,shadows,arrows} 
\usepackage{jheppubmod}

	\def\nn{\nonumber}

	\def\herm{^{\dag}}
	\def\imply{\Longrightarrow}
	
	\def\eq {\equiv}
	\def\l{ \left }
	\def\r{ \right }
	
	\def\p{\partial}
	\def\d { \text{d} }				
	\def\be{\begin{equation}}
	\def\ee{\end{equation}}
	\def\ba{\begin{eqnarray}}
	\def\ea{\end{eqnarray}}
	\def\bpmat{\begin{pmatrix}}  
	\def\epmat{\end{pmatrix}}
	\def\bstep{\left\{ \begin{array}{cc} } 
	\def\estep{\end{array} }
	\def\bmat{\begin{matrix}}				
	\def\emat{\end{matrix}}

	\def\mpl{M_{\rm Pl}}
	\def\stu{St\"uckelberg }
	\def\poin{Poincar\'{e} }

	\def\ket{\rangle}
	
	\def\sla{\slashed}
	\def\gfive{\gamma_5}

	\def\al{\alpha}
	\def\ga{\gamma}
	\def\Ga{\Gamma}
	\def\de{\delta}
	\def\De{\Delta}
	
 \def\ep{\varepsilon}
 \def\vep{\epsilon}
 
 \def\th{\theta}
 \def\la{\lambda}
 \def\La{\Lambda}
 \def\si{\sigma}
 \def\ph{\varphi}
 \def\Ph{\Phi}
 \def\om{\omega}
 \def\Om{\Omega}

	\def\C{\mathcal{C}}

	\def\F{\mathcal{F}}
	\def\G{\mathcal{G}}

	\def\L{\mathcal{L}}
	
	\def\N{\mathcal{N}}
	\def\O{\mathcal{O}}
	\def\P{\mathcal{P}}
	
	\def\R{\mathcal{R}}
	\def\S{\mathcal{S}}
	\def\T{\mathcal{T}}

	\def\spin{\mathfrak{spin}}

	\def\rtwo{\sqrt{2}}
	\def\rthree{\sqrt{3}}
	\def\rsix{\sqrt{6}}
	\def\rtwothree{\sqrt{\frac{2}{3}}}
	\def\rthreetwo{\sqrt{\frac{3}{2}}}
	\def\ronesix{\frac{1}{\sqrt{6}}}
	\def\ronetwo{\frac{1}{\sqrt{2}}}
	\def\risix{\frac{i}{\sqrt{6}}}
	\def\ohalf{\frac{1}{2}}
	\def\thalf{\frac{3}{2}}

	\def\bvep{\bar{\epsilon} }
	\def\bep{\bar{\varepsilon} }
	\def\bpsi{\bar{\psi}}
	\def\bPsi{\bar{\Psi}}

	\def\bchi{\bar{\chi}}
	\def\bQ{\bar{Q}}
	\def\tilh{\tilde{h}}
	\def\tilpsi{\tilde{\psi}}
	\def\tilA{\tilde{A}}

\begin{document}
\title{Deconstructing Supergravity: \\ Massive Supermultiplets}

\author{Nicholas A. Ondo$^{a,b}$, Andrew J. Tolley$^{a,b}$}
\affiliation{$^{a}$Blackett Laboratory, Imperial College, London, SW7 2AZ, U.K.}
\affiliation{$^b$CERCA/Department of Physics, Case Western Reserve University, 10900 Euclid Ave, Cleveland, OH 44106, USA}
\emailAdd{n.ondo16@imperial.ac.uk}
\emailAdd{a.tolley@imperial.ac.uk}


\abstract{

Given the success of the deconstruction program in obtaining ghost-free massive gravity from 5-D Einstein gravity, we propose a modification of the deconstruction procedure that incorporates supersymmetry at the linear level.  We discuss the relevant limits of a conjectured interacting theory of a massive spin 2 supermultiplet, and determine the linear theory to be the $\N=1$ Zinoviev theory, a supersymmetric extension of Fierz-Pauli theory. We develop a family of 1-site deconstruction procedures for fermionic fields (yielding Dirac and Majorana mass terms).  The deconstruction procedure appropriate for giving fermions a Dirac mass is found to preserve half of the supersymmetry of the 5-D theory.  We explicitly check this by deconstructing 5-D $\N=2$ super-Maxwell theory down to 4-D $\N=1$ super-Proca theory, and deconstructing linear 5-D $\N=2$ supergravity down to 4-D $\N=1$ Zinoviev theory, and derive the full 4-D supersymmetry algebras and \stu symmetries from the 5-D superalgebras and gauge symmetries, respectively.  We conjecture that this procedure should admit a generalization to fully non-linear theories.

}

\maketitle


\section{Introduction}
\label{sec:sec1}

Recent advances in the understanding of massive spin-$2$ fields has lead to a renewed interest in their study \cite{deRham:2010gu, deRham:2010ik, deRham:2010kj,Hassan:2011hr} (for a recent review, see \cite{deRham:2014zqa}).  It is now well-understood that there exists a tree-level, ghost-free Lagrangian description of a Lorentz-invariant, self-interacting, massive spin-$2$ field.  This Lagrangian has connections to exotic scalar field theories, known as Galileons, that modify the intermediate IR behavior of the theory \cite{Nicolis:2008in}.  There are many interesting features of massive spin-$2$ fields, but it is not yet known if they are consistent above their intrinsic strong-coupling scale.  Methods for addressing the quantum consistency of these theories, if possible, are only just beginning to be developed \cite{deRham:2012ew, deRham:2013qqa, Cheung:2016yqr,Keltner:2015xda}, and it is an open question of whether these theories admit a standard local UV completion, or some alternative. For the purposes of this paper, we will remain agnostic to their UV consistency, and work at the level of the consistency of the low energy effective theories.

A standard question in field theory is to ask whether it is possible to enhance the symmetries of the theory to include supersymmetry.  Since massive gravity is an interacting theory of a single massless spin 2 field, this amounts to asking whether it is possible to have an interacting theory of a single supermultiplet containing a single spin 2 field.
Incorporating supersymmetry into physical theories has a history of leading to deep insights, many of which seem difficult to discover any other way. The quantum structure of supersymmetric theories are simplified because quantum corrections are required to obey the non-anomalous symmetries of the action.  While internal symmetries reduce some complexity, the true simplifications come from the spacetime symmetries from which the representations of particles descend (e.g. the massive spin-2 representation).  
Enhancing massive gravity to include supersymmetry may provide some insight into its quantum consistency.

This paper contributes to this program by identifying the correct linearized massive supermultiplet, and expanding the dimensional deconstruction program to preserve a single 4-D supercharge from a 5-D theory. In the introduction, we review the salient features of supersymmetry and massive gravity.  In the second section, we discuss our conjectured non-linear theory and explore constraints coming from SUSY and the Vainshtein philosophy, and derive that the Zinoviev theory of a massive superspin-$\thalf$ multiplet is the correct candidate linear theory.  We conclude the second section by reviewing the Zinoviev theory.  In the third section, we develop an extension of the deconstruction program that works to produce the correct fermion mass terms fpr $\N=1$ SUSY. As a proof of principle, we use deconstruction to generate the $D=4$, $\N=1$ super-Proca theory from $D=5$, $\N=2$ super-Maxwell theory, whose supercharge explicitly descends from the higher dimensional supercharge.  In the fourth section, we use deconstruction on the case of linearized $D=5$, $\N=2$ supergravity to obtain the $\N=1$ Zinoviev theory.  This sets up a concrete method for deriving candidate non-linear Lagrangians for the self-interacting, supersymmetric spin-$2$ fields, but we leave their explicit construction for future work. Finally let us comment on the relation of this work to earlier approaches. References \cite{Malaeb:2013lia} and \cite{Malaeb:2013nra} give a proposal for a supersymmetric version of massive gravity. However, the bosonic theory is not the one considered here, but rather that of \cite{Chamseddine:2010ub} which nonlinearly contains ghosts. Reference \cite{DelMonte:2016czb} gives a superfield generalization of the ghost-free massive gravity Lagrangian. While this maintains local sypersymmetry, it does not necessarily ensure a vacuum with global supersymmetry, which is necessary to be viewed as an interacting theory of a single spin 2 supermultiplet.  A more recent work uses a constructive point of view to attempt to construct the leading interactions for the supermultiplet considered here \cite{Zinoviev:2018juc}. In closely related work \cite{Garcia-Saenz:2018wnw} performs an analysis of the supersymmetric partially massless spin states that arise on anti-de Sitter which are effectively special limits of the massive spin supermultiplets with enhanced symmetries.

\subsection{Massive Spin-2 Fields and the Vainshtein Philosophy}

Massive spin-2 fields have a long history in physics, with the linear theory going back to the 1939 paper by Fierz and Pauli \cite{Fierz:1939ix,Fierz:1939zz}.  This theory consists of supplementing the linearized Einstein-Hilbert action with a specific, tuned mass term:
\be
	\L_{\text{mass}} = \frac{1}{2}m^2 \Big( h_{\mu\nu} h^{\mu\nu} - (h_{\mu}\,^{\mu})^2 \Big) = \frac{1}{2} m^2 \de^{\mu\nu}_{\al\beta} h_{\mu}\,^{\al} h_{\nu}\,^\beta \, , 
\ee
where in the last equality we have employed generalized Kronecker delta notation (see Appendix \ref{appsec:usefulform} for details).  This specific combination of the mass terms comes from imposing the absence of ghosts in the linear theory, which is easiest to see in generalized Kronecker delta form because the crucial anti-symmetry is made manifest.  In accordance with Wigner's classification, this theory propagates 5 healthy propagating degrees of freedom, which one may juxtapose to the 2 propagating degrees of freedom of GR.

It was discovered later in the 70's that there was a peculiar feature of this theory, namely that if one used massive gravitons to propagate a gravitational force between two sources, one discovers an enhanced gravitational force even in the limit that $m\to 0$ \cite{vanDam:1970vg,Zakharov:1970cc}.  This can be seen, for instance, if one computes the graviton exchange amplitude from a point source of mass $M$ to generate a potential for a test mass.  One finds two distinct gravitational potentials for linearized GR and for massive gravity as $m\to 0$:
\ba
V_{\text{FP}}(r) &=& - \frac{4}{3} \frac{M}{\mpl^2} \frac{1}{8\pi r}  \\
V_{\text{Lin EH}}(r) &=& - \frac{M}{\mpl^2} \frac{1}{8\pi r}
\ea
This discontinuity exists because of the extra degrees of freedom of the massive spin-2 field, which do not disappear in the limit $m\to 0$.  The five degrees of freedom can be decomposed into helicity states: two helicity-$2$ modes that cause their normal amount of gravitational force, two helicity-$1$ states that decouple from the point source, and finally a single helicity-0 mode that is directly responsible to the additional exchanged force between the two point sources.  From this, one might guess that massive gravity could never be relevant to the real world because of solar system tests implications of this fifth force.

The resolution --which comes for free-- to this problem was noted by Vainshtein: This analysis is linear, and does not take into account the self-interactions for the massive graviton. Vainshtein noticed that adding any self-interactions for the graviton would cause the linear theory to never be valid in the limit that $r \ll m^{-1}$ (Using natural units, $\hbar = c =1$).  The specific breakdown occurs at a scale, called the Vainshtein radius, which is a theory-dependent function of the graviton mass, $\mpl$, and the mass of the point source.  This perspective, that non-linear, strong-coupling physics comes into the theory to restore continuity with (linearized) General Relativity forms the basis of the modern perspective on massive gravity \cite{Vainshtein:1972sx} (for a recent review, see \cite{Babichev:2013usa,deRham:2014zqa}). 

It was pointed out by Boulware and Deser, however, that one is not allowed to add in generic mass terms and self-interactions to Fierz-Pauli without reintroducing the ghostly 6th degree of freedom even if it was removed by the Fierz-Pauli tuning at linear level \cite{Boulware:1973my} (also see related issues \cite{Creminelli:2005qk,Aragone:1971kh,Aragone:1979bm}).  This does not preclude, however, that a specific set of mass terms with a non-linear analog of the Fierz-Pauli tuning to remove the non-linear Boulware-Deser ghost.  Such a theory for a self-interacting, ghost-free massive spin-2 field was recently discovered in \cite{deRham:2010ik,deRham:2010kj} and robustly proven to be free of the BD ghost \cite{Hassan:2011ea,Hassan:2011tf, Mirbabayi:2011aa}.

\subsection{Deconstruction and Ghost-Free, Self-Interacting Massive Spin-2 Fields}

The dimensional deconstruction program \cite{ArkaniHamed:2001nc,ArkaniHamed:2002sp,Schwartz:2003vj,ArkaniHamed:2003vb} exploits the expectation that lower dimensional massless and massive representations can be made to sit inside a higher dimensional representation, and is a procedure for
deforming higher dimensional theories into lower dimensional massive theories according to the decomposition of the representations. 

 The procedure is very similar to 5-D Kaluza-Klein compactification \cite{Chamseddine:1980sp,Dolan:1983aa,Dolan:1984fm}, and represents a specific kind of dimensional reduction (although the resulting theory has no higher dimensional interpretation).  In dimensional reduction, there is an extra dimension, $y$, isolated from the 4 other directions $x^{\mu}$ which has an inverse length scale $M$.  Compactification proceeds by treating the derivatives/integrals of a 5-D field $\Phi$ involving the fifth dimension through Fourier transformations:
\ba
	\Phi(x,y) &= &  \sum_{n=-\infty}^{\infty}{ e^{i 2 \pi nyM} \Phi_n(x)} \, , \nn \\
	\p_y \Phi(x,y) &= & M \sum_{n=-\infty}^{\infty}{ i 2 \pi n e^{inyM} \Phi_n(x)}  \, ,\nn \\
	\int \d y \, \Phi(x,y)\Psi(x,y) &= & \frac{1}{M} \sum_{n=-\infty}^{\infty} \Phi_n(x)\Psi_{-n}(x) \, .
\ea
Instead of keeping the extra dimension as a physical dimension, deconstruction outright deforms the derivative in the extra dimension to some linear operator in some new `site' basis, and abandons the interpretation of being a higher dimensional theory.  Thus, the deformation is a substitution of the form \cite{ArkaniHamed:2001nc,Deffayet:2003zk,Deffayet:2004ws,Deffayet:2005yn,deRham:2013awa}
\ba
	\Phi(x,y) &\to & \Phi_I(x) \, ,\nn \\
	\p_y \Phi(x,y) &\to & M \sum_{J = 1}^{N}{\De_{IJ} \Phi_J(x)} \, ,\nn \\
	\int \d y \, \Phi(x,y)\Psi(x,y) &\to & \frac{1}{M} \sum_{I=1}^{N} \Phi_I(x)\Psi_I(x) \, ,
\ea
where the null eigenspace of $\De$ represents the massless modes and the non-null eigenspace yields the massive modes where the eigenvalues yield the mass of the modes. In other words deconstruction considers the 5th dimension to be a lattice. The deconstructed theory is then fundamentally a set of interacting 4 dimensional theories that may retain a number of properties of its five dimensional starting point. 
The price one generically pays, is the lack of a Leibniz rule and an integration by parts identity, which could break nice or necessary properties of the higher dimensional theory --usually symmetries.  The most commonly used deconstructed derivative is simply, $\De \Phi_1 \sim M (\Phi_2 - \Phi_1)$.  This will be discussed further in section \ref{sec:sec3}, but for now we take it as an ansatz.  

It was recently shown that the ghost-free, self-interacting theory of massive gravity may be obtained via the deconstruction procedure applied to 5-D General Relativity \cite{deRham:2013awa}.  We briefly review this here. Consider the special case of 5-D pure gravity to 4-D massive gravity (or generically multi-gravity).  Pure 5-D General Relativity, here in the Einstein-Cartan formulation has the following action\footnote{Here we are using p-forms where the wedge product is implied, i.e. $A\wedge B \equiv AB$.}:
\be
	\S_{\text{EC}}[E,\Om] = \frac{M^3_5}{6} \int \ep_{ABCDE}\R^{AB} E^C E^D E^E \, ,
\ee
where $E^A$ is the vielbein (f\"unfbein) and $\Om^{AB}$ is the 5-D spin connection, which sits inside the Riemann tensor as
\be
	\R^{AB} = d\Om^{AB} + \Om^{A}\,_{\bullet}\Om^{\bullet B} \,\, .
\ee
Upon integrating out the auxiliary field $\Om^{AB}$, we obtain the torsion condition
\be
	dE^A + \Om^{A}\,_B E^B = 0 \,\, . \label{eqn:tor_cond}
\ee
The simplest form of deconstruction proceeds by exhausting all gauge symmetries; we gauge fix\footnote{This gauge choice assumes that the proper size of the extra dimension (the radion) is fixed. Turning on the radion would correspond to the less restrictive choice $E_y^4=e^{\phi(x)}$. This is an additional global constraint that appears to be an essential part of the deconstruction procedure. Beyond this, the remaining gauge choices may always be chosen locally, and for a sufficiently small extra dimension would also be expected to be valid globally. It remains an open question whether the deconstruction procedure can be applied successfully with a less restrictive gauge choice.} such that
\ba
\begin{matrix}
	E_{\mu}\,^a &=& e_{\mu}\,^a &\, , \,\,\,\,\,\,\,\,& 	E_y\,^a &=& E_{\mu}\,^4 & = 0 \, , \\
	E_y \,^4 &=& 1  & \, , \,\,\,\,\,\,\,\, &	\Om_y\,^{ab} &=& 0 \, ,
\end{matrix}
\ea
which when substituted into (\ref{eqn:tor_cond}), imposes the following conditions on the spin connection
\ba
	\Om_y \,^{a4} &=&  0 \, ,\\
	\Om_{\mu}\,^{a4} &=& K_{\mu}\,^a = \p_y e_{\mu}\,^a \, , \\
	\Om_{\mu}\,^{ab} &=& \om_{\mu}\,^{ab} \, ,
\ea
where in the last line, $\om_{\mu}\,^{ab}$ is the usual 4-D spin connection, and is a function of the 4-D vierbein in the usual manner.  Substituting this into the action, we obtain
\ba
	\S_{\text{Dec}} = \frac{M^3_5}{4} \int \d y \int \l( \R^{ab}[\om] + m^2 K^a K^b \r) e^c e^d  \ep_{abcd}\, .
\ea
Applying a simple 2-site model, i.e. two vielbein $e^a$ and $f^a$, with the discretization operator as the replacement for the $y$ derivative, which is
\be
	\p_y e^a \to m  \Big( e^a - f^a \Big) \, \, .
\ee
and taking $f^a$ to be non-dynamically (by sending its Planck mass to infinity), so that $f^a = \delta^a$ is the Minkowski vacuum, we obtain the ghost-free, self-interacting massive gravity:
\ba
	\S_{\text{dRGT}} = \frac{\mpl^2}{4} \int \left[ \R^{ab}e^c e^d + m^2 \Big( e^a - f^a \Big)\Big( e^b - f^b \Big) e^c e^d \right] \ep_{abcd}\,\, ,
\ea
subject to a constraint arising from gauge-fixing
\be
	\Om_y\,^{ab} = K^a e^b =0 \Rightarrow e^a f_a = 0 \, ,
\ee
which is the Deser-van Nieuwenhuizen condition (aka ``symmetric vielbein'' condition) \cite{ Hinterbichler:2012cn,Deffayet:2012zc, Deser:1974cy}.  For a full discussion of how to restore diffeomorphisms and local Lorentz symmetries and how to obtain to the metric formulation of the theory, see \cite{Ondo:2013wka,Hassan:2012wt}.

This demonstrates cleanly that a ghost-free, non-linear extension of massive gravity can be obtained from deforming an ordinary 5-D Einstein gravity theory, which suggests a surprising relationship between ghost-free massive gravity and GR.  This surprising correspondence suggests that a non-linear theory of self-interacting, massive supermultiplet may be obtainable from a similar procedure.


\section{Survey of Massive Supermultiplets Containing Spin-2 Fields}
\label{sec:sec2}

\subsection{Supersymmetry and Supergravity}

We begin this section with a brief discussion of supergravity; note that our conventions for fermions are given in the appendix \ref{sec:appA}.  It is well-known from the Coleman-Mandula theorem that the kinds of linearly-realized symmetries present in a QFT are very restricted. Although standard QFT only makes use of the \poin group and internal groups, the symmetries of field theories may be extended to include fermionic generators (supercharges), $Q^i$,
 and R-symmetry charges $R$ (which rotate the supercharges).  Together they must obey the super-\poin algebra:
\ba
	&& \, \{ Q^i, \bQ^j \}= 2i \gamma^a P_a \, \de^{ij} \, , \\
	&& \, [ Q^i , R ]= i\gfive Q^i \, ,
\ea
in addition to the usual \poin commutation relations and all other (anti-)commutation relations vanishing (for recent references on supersymmetry and supergravity, see \cite{Quevedo:2010ui,Tanii:1998px,deWit:2002vz}). It is well-known that GR is the unique interacting theory of a massless spin-$2$ field \cite{Gupta:1954zz,Feynman:1996kb,Weinberg:1965rz,Deser:1969wk,Boulware:1974sr}, and in Einstein-Cartan formulation the vielbein and spin connection can be viewed as the gauge fields for a local \poin group \cite{Kiriushcheva:2009tg,Kiriushcheva:2009nj}. Unlike normal gauge theories, the Einstein-Hilbert action is manifestly invariant under the local Lorentz group, but realizes the translations as diffeomorphisms, rather than as standard gauge transformations.

 In the same way `supergravity' theories may be viewed as interacting theories of massless supermultiplets, or gauge theories of a local super-\poin group. The local super-\poin group has fermionic generators, which form the gauge redundancies for the massless spin-$\thalf$ fields, and the remaining bosonic generators form the gauge redundencies of its massless bosonic superpartners \cite{Freedman:1976xh,Deser:1976eh,Grisaru:1976vm}.   To linear order, the action for supergravity is given by the kinetic terms for the graviton, $h_{\mu\nu}$ and it's superpartner the gravitino, $\psi_{\mu}$, i.e.
\ba
	\S [h, \psi] = \int \d^4x \left[  -\frac{1}{2} h_{\mu}\,^{\al} \de^{\mu \nu \rho}_{\al \beta \ga} \p_{\nu} \p^{\beta} h_{\rho}\,^{\ga} - \frac{i}{2} \bpsi_{\mu}  \ga^{\mu\nu\rho} \, \p_{\nu} \psi_{\rho} \right] \, .
\ea
This is invariant under $\N=1$ supersymmetry transformations
\ba
&&	\de h_{\mu\nu} =i  \bvep \ga_{(\mu} \psi_{\nu)}  \, ,  \nn \\
&&	\de \psi_{\mu} = \ga^{ab} \p_{a}h_{b\mu} \vep \, .
\ea
There are some immediate structural changes to the Lagrangian for the supermultiplet  if the graviton is given a mass, most notably in the form of change to the superspin-$Y$ representations of the super-\poin algebra.  While in four dimensions, massless representations always come in pairs of 2, the massive representations always come in pairs of 4, i.e.
\ba
		\bpmat & Y+ \ohalf & \\ Y & & \hat{Y} \\ & Y- \ohalf  & \epmat \, ,
\ea
because there are twice as many helicity mixing operators in the supercharge (for reviews, see \cite{deWit:2002vz} and \cite{Quevedo:2010ui}).  One operator increments and the other decrements the helicity by $1/2$ to build the full super representation, and they transform oppositely under the axial $U(1)$ R-symmetry.  Thus, $Y$ and $\hat{Y}$ transform oppositely under R-symmetry.


\subsection{A Survey of the Candidate $\N =1$ Supermultiplets}

There are precisely three supermultiplets that contain a massive spin-2 field
\ba
		\bpmat & 2 & \\ 3/2 & & 3/2 \\ & 1  & \epmat \, , \, \bpmat & 5/2 & \\ 2 & & 2 \\ & 3/2  & \epmat \, , \,  \bpmat & 3 & \\ 5/2 & & 5/2 \\ & 2  & \epmat \, ,
\ea
There is only one supermultiplet which does not contain higher spin fields and thus will not require higher spin interactions, which is the superspin-$\thalf$ multiplet containing $(2, \, 3/2, \, 3/2 \, , 1 )$.  Therefore, assuming that we do not want an infinite tower, we are looking for the largest spin being 2. Thus we are lead to conjecturing a theory with one massive spin-2 field, two massive Majorana spin-$\thalf$ fields, and one massive spin-1 field.  The dof counting then goes $4+4 = 5+3$, since a massive Majorana spin-$\thalf$ has 4 dof, a massive spin-2 has 5 dof, and a massive spin-1 has 3 dof in four dimensions. The question we would like to address is whether there is a consistent interacting theory of a single superspin-$\thalf$ multiplet.

\subsection{Properties of the Conjectured Fully Non-Linear Theory}
\label{subsec:props}

Although we shall only be concerned here with the linear theory, looking ahead one may understand a large amount about the conjectured non-linear theory merely by analyzing various scaling limits of the theory, and derive consistency conditions from this.  The theory without matter couplings is expected to have only two scales, the first being the gravitational coupling constant $\mpl$ and the second scale being the mass of the graviton itself $m$.  
The Vainshtein interactions prohibit a direct scaling $m \to 0$, since this limit is divergent due to interactions in the Lagrangian scaling as
\be
	\l(\frac{1}{\mpl m^2}\r)^n \O^{4+3n}  \, .
\ee
Thus from the perspective of the bosonic theory, there are only two limits of the theory that we have good reason to believe that we understand, illustrated in figure (\ref{fig:scalingdiagram}).  The first is the decoupling limit:
\ba
&&	m  \to  0 \nn \, , \quad	\mpl  \to  \infty \nn \, ,\\
&&	\mpl m^2  =  \La^3_3 \,, \quad finite \, .
\ea
In the case of pure massive gravity, this limit leads to a massless spin 2 field decoupled from an interacting theory of a massless vector field and a massless scalar Galileon theory \cite{Nicolis:2008in,deRham:2010gu,deRham:2010ik,deRham:2010kj,Ondo:2013wka,Gabadadze:2013ria}.  In the supersymmetric case, we expect a similar limit to exist.  Since a massive spin-2 supermuliplet decomposes in the massless limit into one massless $\N=2$ spin-$2$ supermultiplet, one massless spin-$1$ $\N=2$ supermultiplet; we expect that at least one of the spin-0 and spin-1 fields that arise in the later to have `Galileon' interactions similar to the non-supersymmetric case. The pair of spin-$\thalf$ fields, being massless in this limit, must propagate two full local supersymmetries, which gives rise to supersymmetry uplifting in the $m\to 0$ limit.  We anticipate that as in the non supersymmetric case, the massless spin-$2$ supermultiplet decouples, and simply gives a copy of linearized $\N=2$ supergravity. The remaining vector-scalar theory may be viewed as an $\N=2$ Supersymmetric extension of a Galileon theory (coupled to a Maxwell field).
This will be an interesting limit of the non-linear theory, but until there is a method to construct the fully non-linear theory and take this limit, not much else can be said, thus we move onto the second limit.

The second important limit is the where we drop all gravitational self-interactions:
\be
	\mpl \to \infty \, ,
\ee
but we keep the mass as a free parameter. This limit simply linearizes the theory, and we will be led to a linear $\N=1$ supersymmetric theory of a massive spin 2 supermultiplet whose form will be outlined below.

\vspace{5pt}
	\tikzstyle{line} = [draw, -stealth, thick]
	\tikzstyle{bubble1} = [draw, rectangle, fill=blue!35, text width=9em, text centered, minimum height=15mm, node distance=10em]
	\tikzstyle{bubble2} = [draw, rectangle, fill=blue!35, text width=13em, text centered, minimum height=15mm, node distance=10em]
	\tikzstyle{limit} = [draw, minimum height=8mm, text width=6em, text centered]
	\begin{tikzpicture}
		\node [bubble1] (main) {Interacting Massive superspin-$\thalf$ multiplet};
		\node [bubble2, above of=main, xshift=14em, yshift=2em] (top) {$\N=2 \text{ Linear SUGRA}+$ $\N=2 \text{ SUSY Galileon}$ };
		\node [bubble2, below of=main, xshift=14em, yshift=-2em] (bottom) {$N=1$ Zinoviev};
		\node [bubble2, right of=main, xshift=18em] (side) {$\N=2 \text{ Linear SUGRA+}$ $\N=2$ spin-1 supermultiplet};
		\node [limit, above of=main, xshift=4em, yshift=4em] (decoup) { $ m \rightarrow 0 $ $M_{pl}\rightarrow \infty $  $\Lambda_3 =m^2 M_{pl}  $};
		\node [limit, below of=main, xshift=4em, yshift=-3em] (zinoviev) {$ M_{pl} \rightarrow \infty $ };
		\node [limit, above of=bottom, xshift=10em, yshift=3em] (vDVZ) {$ m \rightarrow 0 $ };
		\node [limit, below of=top, xshift=10em, yshift=-3em] (vainshtein) {$ \Lambda_3 \rightarrow \infty$ };
		\path [line] (main) -- (decoup);
		\path [line] (decoup) -- (top);
		\path [line] (main) -- (zinoviev);
		\path [line] (zinoviev) -- (bottom);
		\path [line] (bottom) -- (vDVZ); 
		\path [line] (vDVZ) -- (side);
		\path [line] (top) -- (vainshtein);
		\path [line] (vainshtein) -- (side);
	\end{tikzpicture}
\vspace{-20pt}
\begin{figure}[h!]
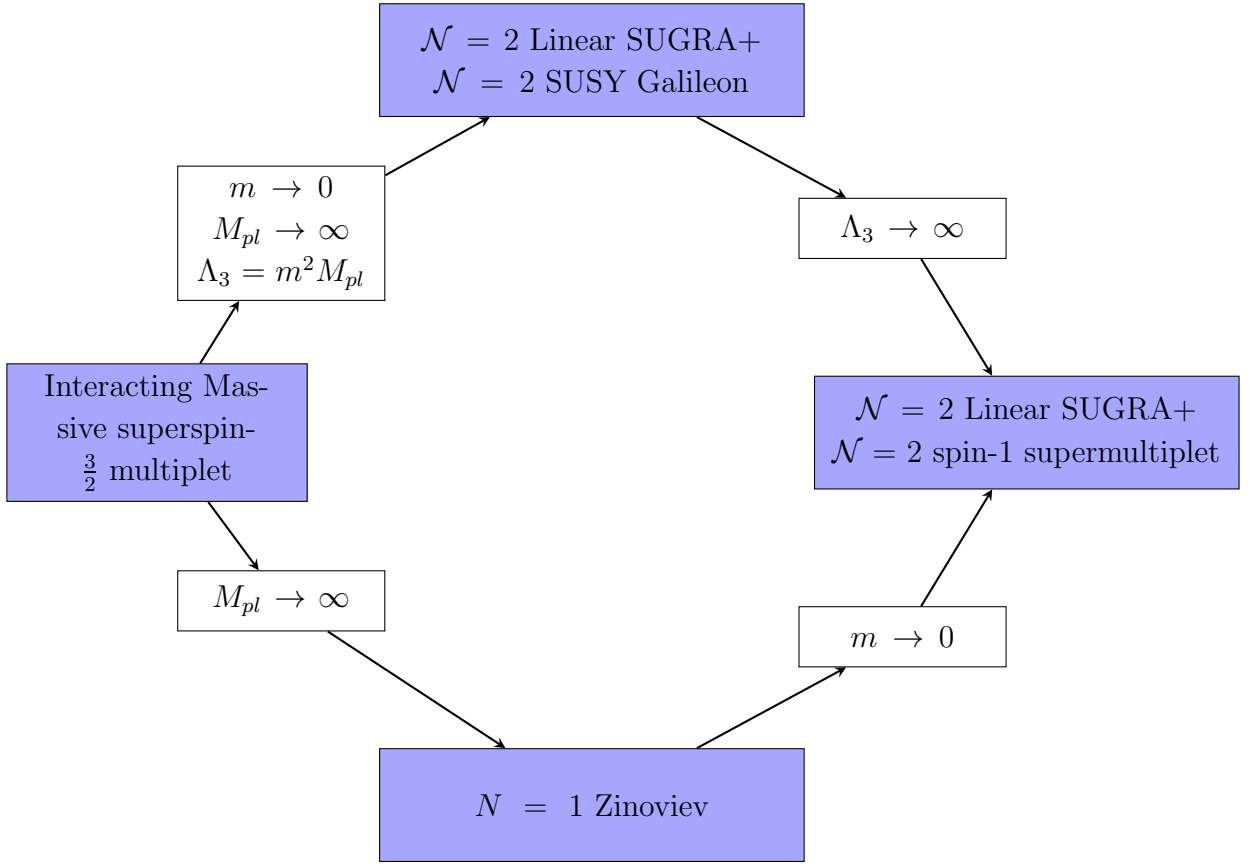

	\begin{center}
		\caption{\small{The scaling limits of the conjectured interacting theory of a massive supermultiplet. Notice that each limit of the theory ends (when $m=0$) with $\N=2$ SUSY, even though the original massive theory only has $\N=1$.   }}
	\end{center}
\label{fig:scalingdiagram}	
\end{figure}

In this paper, we will construct a method for deriving the $\N=1$ linearized massive theory through deconstruction, which we hope to be generalizable to the non-linear theory.  Before finishing off this section, we will mention an important point regarding the fully non-linear theory, which is that it has a pair of spontaneously broken supersymmetries.  To take the massless limit, these local symmetries must be reintroduced as \stu fields. The situation is an extension of the fact that massive gravity on Minkowski respects a global Poincare symmetry but breaks local diffeomorphisms. In the massive case, to make the $\N=1$ symmetry local we will need to introduce an additional vielbein and gravitino which act as the gauge fields for this symmetry. This will lead to a locally supersymmetric theory of bigravity describing a massive supermultiplet interacting with a massless supermultiplet. In principle this is straightforward to accommodate in the non-linear theory through the deconstruction framework, just as it is know that we can obtain multi-vielbein gravity models from deconstruction \cite{Hinterbichler:2012cn,deRham:2013awa}.
\subsection{The Zinoviev Theory of Massive $\N=1 ,\, Y= \thalf$ Supermultiplets}

The second scaling limit is the linearized limit, i.e. $\mpl\to\infty$.  The appropriate supersymmetric quadratic Lagrangian describing this limit has been given explicitly by Zinoviev \cite{Zinoviev:2002xn}, hence we will refer to it as the $\N=1$ Zinoviev theory. 
Since we expect to have a supermultiplet containing 1 massive spin-$2$ field, 2 massive spin-$\thalf$ fields, and 1 massive spin-$1$ field, several key things must be specified in the linear theory:
\begin{itemize}
	\item[1.) ] The fermions can have different kinds of mass terms, i.e. Dirac or Majorana.
	\item[2.) ] SUSY mandates that certain spins in the supermultiplet must be $\P\T$-odd.
\end{itemize}
We will address these issues in reverse order. Although there are group-theory based arguments for the $\P\T$-charge assignments in the supermultiplet, we will make a simple observation which will gives the correct answer. The observation is to perform a helicity decomposition on the massive spin states (i.e. the decoupling limit), noting that all of the helicities from the massive state must share the same $\P\T$-charge assignment. The massive spin-2 field $\tilh_{\mu\nu}$ decomposes into the helicity-$(\pm2)$ states, the helicity-$(\pm1)$ states, and a helicity-$0$ state (i.e. $h_{\mu\nu}, \, B_{\nu}, \, \pi$).  The two massive spin-$\thalf$ fields $\tilpsi_{\mu}\,^i$ decompose into two helicity-$(\pm\thalf)$ states and two separate helicity-$(\pm\ohalf)$ states (i.e. $\psi_{\mu}^i,\, \chi^i$, where $i$ labels the two massive spins).  Finally, the massive spin-$1$ field $\tilA_{\mu}$ decomposes into the helicity-$(\pm1)$ states and a helicity-$0$ state (i.e. $A_{\mu},\,\ph$).

Then in this limit, we know that one of the $\chi^i$ along with $\pi$ and $\ph$ must form a Wess-Zumino multiplet, and therefore one of the scalars must be a pseudo-scalar.  This means that either the massive spin-$2$ state or the massive spin-$1$ state must be $\P\T$-odd.  Supergravity maintains that the helicity-$2$ states are not axial.  Therefore, $A_{\mu}$ must be an axial, massive spin-1 field.

As for the first question, we know that each fermion is oppositely charged under R-symmetry, thus the R-symmetry transformation is
\ba
	\bpmat \psi_{\mu}\,^1 \\ \psi_{\mu}\,^2 \epmat	\rightarrow  \bpmat e^{i\th \gfive} \, \psi_{\mu}\,^1 \\ e^{-i\th \gfive} \, \psi_{\mu}\,^2 \epmat \label{eqn:Rtrans}  \, , \\ \Rightarrow \psi_{\mu}\,^i \rightarrow \l(e^{i \theta \gfive \eta} \r)^i\,_j \, \psi_{\mu}\,^j \, .	
\ea
Then, the task becomes to find mass terms consistent with $R$-symmetry invariance; we find we have only one correct mass term.  Structurally at quadratic order, the only allowed (i.e. a Fierz-Pauli tuning to remove ghosts) form of the mass terms are:
\be
	\int \d^4 x\,  \left( \frac{1}{2} m \bpsi_{\mu}\,^i \ga^{\mu\nu} A^{ij}\psi_{\nu}\,^j \right) \, , 
\ee
where the antisymmetric $\ga^{\mu\nu}$ projects out the ghost.  Of the two potential mass terms, the Majorana mass term and the Dirac mass term, are expressed through different choices of $A^{ij}$\footnote{The other two choices are $\ep^{ij}$ which is identically zero, and $\eta^{ij}$ which is field redefinable to the Dirac mass $\De^{ij}$.}:
\ba
	\S_{\text{Dirac}} = \int \d^4 x \, \left(  \frac{1}{2} \bpsi_{\mu}\,^i \ga^{\mu\nu} \De^{ij} \psi_{\nu}\,^j \label{eqn:diracmass}  \right) \, , \\
	\S_{\text{Majorana}} = \int \d^4 x \, \left( \frac{1}{2} \bpsi_{\mu}\,^i \ga^{\mu\nu} \psi_{\nu}\,^i   \right) \, , 
\ea
where $\De^{ij}$ is the 2x2 matrix described in the appendix.  The $\De^{ij}$ structure makes the action manifestly invariant under (\ref{eqn:Rtrans}), including the kinetic terms.  Thus, we must take the Dirac mass.

\subsubsection{The Zinoviev Lagrangian and \stu Symmetries}

Collecting these results of the previous section, we have the Zinoviev Lagrangian:
\ba
		\S[h , \psi, A] &=& \int \d^4x  \left[ -\frac{1}{2} h_{\mu}\,^{\al} \de^{\mu \nu \rho}_{\al \beta \ga} \p_{\nu} \p^{\beta} h_{\rho}\,^{\ga}  + \frac{1}{2} m^2 h_{\mu}\,^{\al} \de^{\mu\nu}_{\al\beta} h_{\nu}\,^{\beta}     \nn \right. \\
		&&  \,\,- \frac{i}{2} \bpsi_{\mu}\,^i  \ga^{\mu\nu\rho} \, \p_{\nu} \psi_{\rho}\,^i + \frac{1}{2} m \bpsi_{\mu}\,^i \ga^{\mu\nu} \De^{ij} \psi_{\nu}\,^j \nn \\
		&& \left. -\,\, \frac{1}{4} \F_{\mu\nu}\F^{\mu\nu} - \frac{1}{2} m^2 A_{\mu} A^{\mu} \right]  \, , \label{eqn:Zin1Action}
\ea
which is the combined Fierz-Pauli, Rarita-Schwinger-Dirac, and Proca actions, respectively. While this action is a valid formulation of the $\N=1$ Zinoviev Lagrangian, it is much better to introduce all of the \stu symmetries for all of the massive fields which become in the massless limit the usual gauge symmetries of the massless fields.  To do this we introduce the \stu fields, $B_{\mu}, \, \pi, \phi,$ and $\chi^i$ as follows:
\ba
&&	h_{\mu}\,^{\al} \rightarrow h_{\mu}\,^{\al} - \frac{1}{2m} \l(\p_{\mu} B^{\al} + \p^{\al} B_{\mu} \r) + \frac{1}{m^2} \p_{\mu} \p^{\al} \pi \, ,  \\
&&	\psi_{\mu}\,^i \rightarrow \psi_{\mu}\,^i - \frac{1}{m} \p_{\mu} \chi^i \, , \\
&&	\bpsi_{\mu}\,^i \rightarrow \bpsi_{\mu}\,^i - \frac{1}{m} \p_{\mu} \bchi^i \, ,\\
&&	A_{\mu} \rightarrow A_{\mu} - \frac{1}{m} \p_{\mu}\ph \, ,
\ea
which will restore linearized diffeomorphism invariance, the $U(1)$ invariance for the diff vector \stu field, invariance under both of the supergauge symmetries of the two Rarita-Schwinger fields, and finally the $U(1)$ of the axial vector. This leads to the \stu formulation of the Zinoviev action, where here we zoom in to the mass terms (since the kinetic terms are gauge invariant), ignoring total derivatives:
\ba
	\S_{\text{mass}} &=& \int \d^4 x  \left[ \frac{1}{2} m^2 \de^{\mu\nu}_{\al\beta} \Big( h_{\mu}\,^{\al}h_{\nu}\,^{\beta} - \frac{2}{m} \p_{\mu} B^{\al} h_{\nu}\,^{\beta} + \frac{2}{m^2} \p_{\mu} \p^{\al} \pi  h_{\nu}\,^{\beta} -\frac{1}{m^2} \p_{\mu}B_{\nu}\p^{\al}B^{\beta} \Big)  \right. \nn \\
		&&  + \frac{1}{2} m \bpsi_{\mu}\,^i  \ga^{\mu\nu}\De^{ij} \psi_{\nu}\,^j - \bpsi_{\mu}\,^i  \ga^{\mu\nu}\De^{ij} \p_{\nu} \chi^j \nn \\
		&& \left.  - \frac{1}{2} m^2 A_{\mu} A^{\mu} + m A_{\mu} \p^{\mu}\ph - \frac{1}{2} \p_{\mu}\ph \p^{\mu}\ph  \right]\,\, . \label{eqn:diracmixy}
\ea
As it stands the \stu fields do not have canonical kinetic terms (except for $\ph$, which does have a kinetic term), but they obtain kinetic terms via kinetic mixing. To make the kinetic terms manifest we perform the field redefinitions\ba
&&	\pi \to \rtwothree \pi \, , \\
&&	B_{\mu} \to \ronetwo B_{\mu}  \, ,  \\
&&	\chi^i \to \rtwothree \chi^i \, , 
\ea
and then diagonalize the fields as
\ba
&&	h_{\mu}\,^{\al} \rightarrow h_{\mu}\,^{\al} - \frac{1}{\rsix}\pi \de_{\mu}^{\al} \, , \nn \\
&&	\psi_{\mu}\,^i \rightarrow \psi_{\mu}\,^i - \frac{i}{\rsix}\ga_{\mu} \De^{ij}\chi^j \, , \nn\\
&&	\bpsi_{\mu}\,^i \rightarrow \bpsi_{\mu}\,^i + \frac{i}{\rsix} \bchi^j \De^{ji} \ga_{\mu} \,    \,\, . \label{eqn:helunmixy}
\ea
This leads to the canonically normalized Zinoviev action, which we split up order by order into powers of mass $\S=\S_0+m \S_1+m^2 \S_2$:
\ba
	\S_0 &=& \int \d^4x \,\, \left[  -\frac{1}{2} h_{\mu}\,^{\al} \de^{\mu \nu \rho}_{\al \beta \ga} \p_{\nu} \p^{\beta} h_{\rho}\,^{\ga} - \frac{i}{2} \bpsi_{\mu}\,^i  \ga^{\mu\nu\rho} \, \p_{\nu} \psi_{\rho}\,^i  \right. \nn \\
		&& - \frac{1}{4} \F_{\mu\nu}\F^{\mu\nu} - \frac{1}{4} \G_{\mu\nu}\G^{\mu\nu} +\frac{i}{2} \bchi^i \ga^{\mu} \p_{\mu} \chi^i \nn \\
		&&  \left. - \frac{1}{2} \p_{\mu}\pi\p^{\mu} \pi  - \frac{1}{2} \p_{\mu}\ph\p^{\mu} \ph \right] \, ,\label{eqn:zinoviev0}\\
		&& \nn\\
	m\S_1 &=& \int \d^4x \,\, \left[  -m \rtwo \de^{\mu\nu}_{\al\beta} h_{\mu}\,^{\al} \p_{\nu}B^{\beta} + m\rthree \pi \p_{\mu}B^{\mu}  \right.  \, , \nn \\
		&& \left. +\frac{1}{2} m \bpsi_{\mu}\,^i \ga^{\mu\nu} \De^{ij} \psi_{\nu}\,^j + i m\sqrt{\frac{3}{2}}\bpsi_{\mu}\,^i \ga^{\mu} \chi^i + m \bchi^i \De^{ij} \chi^j +  m A_{\mu} \p^{\mu}\ph  \right]\label{eqn:zinoviev1}\\
	&& \nn\\
	m^2\S_2 &=& \int \d^4x \,\left[ - \frac{1}{2} m^2 A_{\mu} A^{\mu} + \frac{1}{2} m^2 h_{\mu}\,^{\al} \de^{\mu\nu}_{\al\beta} h_{\nu}\,^{\beta} + m^2 \l( \pi^2 - \sqrt{\frac{3}{2}}\pi h_{\mu}\,^{\mu} \r) \right] \,  . \label{eqn:zinoviev2}
\ea

		\subsubsection{Symmetries of the Zinoviev Lagrangian}

This leads to an action with many abelian gauge symmetries.  There are four linearized diffeomorphism symmetries (with bosonic group parameter $\xi_{\mu}$),
\be
 \de h_{\mu\nu} = \p_{(\mu} \xi_{\nu)}  \, ,\quad  \de B_{\mu} = m \rtwo \xi_{\mu} \, , \quad \de \pi = 0 \, , \ee
and two supergauge symmetries (fermionic Majorana group parameter $\eta^i$),
\ba
	&& \de \psi_{\mu}\,^i = \p_{\mu} \eta^i + i \frac{m}{2} \ga_{\mu}\De^{ij} \eta^j\nn \, ,  \\
	&& \de \chi^i = m \rthreetwo \eta^i \, ,
\ea
 and 2 $U(1)$ gauge symmetries (bosonic group parameter $\xi$).  The first is $U(1)$ from the vector mode of the spin-2 field
\ba
		&& \de h_{\mu\nu} = \frac{m}{2} \eta_{\mu\nu}\xi  \, , \nn \\
		&& \de B_{\mu} = \p_{\mu} \xi \, , \nn \\ 
		&& \de \pi = m \rthreetwo \xi \, ,\ea
and the last is the original \stu symmetry for the massive axial vector
\be
	 \de A_{\mu} = \p_{\mu}\theta \nn \, , \quad  \de \phi = m \theta \, .
\ee

Finally, there are the crucial global symmetries.  In addition to the \poin symmetries, there is an additional $\N=1$ supersymmetry, given by the following super-transformations
\ba
	\de h_{\mu\nu} &=&  \al^i \, i\bvep \ga_{(\mu} \psi_{\nu)}\,^i \, ,  \nn\\
	&&\nn\\
	\de \psi^i &=& \al^i\, \ga^{\al\beta}\p_{\al} h_{\beta\mu}\vep -\frac{m}{\rtwo} \Big[ \ga_{\mu}\ga^{\al}B_{\al} + i\rthree \gfive A_{\mu} \Big]\al^i \vep \nn \\
	&& -\frac{i}{4\rtwo} \ga^{\al\beta}\ga_{\mu} \Big[ \G_{\al\beta} - \rthree i\gfive \F_{\al\beta}\Big] \beta^i \vep + im \Big[ \ga^{\al}h_{\al\mu} + \ga_{\mu} \pi \Big] \beta^i \vep  \nn \, , \\
	&&  \nn\\
	\de B_{\mu} &=& \beta^i \, \frac{1}{\rtwo} \bvep \psi^i + \al^i \,i \frac{\rthree}{2} \bep \ga_{\mu} \chi^i \nn \,,  \\
	&&\nn\\\de A_{\mu} &=& \beta^i \, \rthreetwo \bvep \gfive \psi_{\mu}\,^i + \alpha^i \ohalf \bvep \ga_{\mu}\gfive \chi^i \nn \, ,  \\
	&&\nn\\
	\de \chi^i &=& -\frac{1}{4} \ga^{\al\beta} \Big[\rthree \G_{\al\beta} + i\gfive \F_{\al\beta}\Big]\vep \al^i \nn  \\
	&&    -i \ga^{\al}\Big[ \p_{\al}\pi +\gfive \p_{\al}\ph \Big] \beta^i \vep + im \ga^{\al} \Big[ \rthree B_{\al} -i\gfive A_{\al} \Big] \beta^i \vep \nn \, , \\
	&&\nn \,  \\
	\de \pi &=& i \beta^i \bvep \chi^i \nn  \, , \\
	&&\nn\\
	\de \ph &=& \beta^i \, i \bvep \gfive \chi^i  \, , \label{eqn:zinovievtrans}
\ea
where
\ba
	\al^i &=& \bpmat 0 \\ 1 \epmat = -\eta^{ij} \al^j \, ,  \\
	\beta^i &=& \bpmat -1 \\ 0 \epmat = \ep^{ij}\al^j = -\De^{ij} \al^j  \, . \label{eqn:alphaidentities}
\ea
The $i$ indices are a product of $\N=2$ R-symmetry, and thus since we work with an $\N=1$ theory, this must be broken.  The kinetic terms are invariant under arbitrary $\al^i$, with $\beta^i$ still subject to $\beta^i =-\ep^{ij}\al^j$.  But once the mass terms are added, the $\N=2$ R-symmetry is broken down to the $\N=1$ R-symmetry.
  This can be seen at the level of the Lagrangian, where the transformations (\ref{eqn:zinovievtrans}) on the action only cancel the mass terms if the conditions (\ref{eqn:alphaidentities}) are applied. Thus, as expected, when the \stu fields are included, the kinetic terms have an $\N=2$ structure, but the mass terms break half of the SUSY down to an $\N=1$ algebra.

Now that we know what theory needs to be reproduced by the deconstruction of the massless 5-D theory, we now turn to the issue of how to extend the deconstruction procedure to include fermions.

\section{Deconstructing Fermions}
\label{sec:sec3}

For the purposes of this paper, we restrict our interest to 1-site deconstruction, since we are only interested in having a single massive mode and this is the simplest form of dimensional deconstruction with a single mode.  As stated in the introduction, the deconstruction procedure works as follows
\ba
	\Phi(x,y) &\to& \phi(x) \, ,  \\
	\p_y\Phi(x,y) &\to& m \phi(x) \, , \\
	\int \d y \Phi(x,y) \Psi(x,y) &\to& \frac{1}{m}\phi(x) \psi(x) \, . 
\ea
Note that there exist several variants of the deconstruction procedure, based upon how one wishes to discretize the extra dimension $y$.  The correct choice depends upon what one wants to use the deconstruction procedure for. The discretization operator is often chosen to reflect a discretized compact dimension, in which case it reflects a truncated Kaluza-Klein tower, as has been described in the literature before, and has been generalized in a number of ways \cite{Gregoire:2004ic,Nojiri:2004jm,Craig:2014fka}.  None of these theories, however, result in the $\N=1$ Zinoviev theory.

	\subsection{Group-Theoretic Interpretation of Deconstruction}
\label{sec:sec:3.1}
We now quickly review the bosonic deconstruction procedure.  While most efforts in the deconstruction program have emphasized the geometrical interpretation of deconstruction, our interest will largely lie in the group-theoretic interpretation. Thus, the purpose of deconstruction will be to explicate the massive $D=d$ spin-$J$ subrepresentation of the massless $D=(d+1)$ helicity-$J$ representation.  This will generate an ans\"atz for a candidate ghost-free Lagrangian in one lower dimension.

To illustrate schematically how 1-site deconstruction works, let us analyze the deconstruction procedure for the trivial case of the $(d+1)$-dimensional Lagrangian of a free bosonic field, $\Phi$ and its second-order differential operator $\Box$.  Although this looks exclusively tailored for scalar fields, this argument will follow generically for all non-zero spin-$J$ bosonic fields as well, since the only mathematical property needed is that the differential operator is symmetric, i.e. $A \Box B = B \Box A$ in the integrand.  Deconstruction proceeds as follows, beginning with the $(d+1)$-split of the action:
\ba
		\S_{d+1} &=& \int \d^{(d+1)}x \, \, \l( \ohalf \Ph\, \Box_{d+1} \Ph \r) \, , \\
		&=& \int \d^{d}x \,\d y \, \, \l(  \ohalf \Ph \Box \Ph  - \ohalf \p_y \Ph \O^2 \p_y \Ph \r) \, .
\ea
Here the operator $\O^2$ holds the tensor indices (for scalars, it is $1$), and because of the higher dimensional structure, the Lagrangian is ghost-free (for the same essential reasons that compactification produces ghost-free Lagrangians).  Although in general, the spin-$J$ field contains components in the $y$ direction, we may use the gauge freedoms of the massless theory to set these to zero. Then when one performs the $1$-site deconstruction, we see that we explicitly yields the mass terms
\ba
		\S_{\text{Dec}} &=& \int \d^d x \,  \ohalf \ph \Box \ph - \frac{1}{2}m^2 \ph \O^2 \ph  \, . 
\ea
where $\ph$ is a four dimensional tensor constructed from only the four dimensional components of $\Ph$. Thus, from one spin-$J$ field in $(d+1)$-dimensions, we have extracted a massive spin-$J$ field in $d$ dimensions.  It can be verified that this creates a massive Klein-Gordon, Proca, and Fierz-Pauli from one higher dimensional massless Klein-Gordon, Maxwell, and linearized Einstein-Hilbert Lagrangians, respectively.  If one gauge-fixes first, we obtain the above so-called unitary gauge Lagrangian. If we do not gauge-fix, there will be additional fields arising from the various $y$-components of the original spin-$J$ field, and these additional fields will have the interpretation as the \stu fields in the \stu formulation of the massive theory. The \stu formulation is defined as the one on which the massless theory exhibits the same symmetries as the massless one.

	\subsection{Fermionic Deconstruction}

We now restrict our interest to deconstructing massless 5--D fermion theories into massive 4--D fermion theories; as one would expect, this case will be more subtle than the bosonic deconstruction procedure. There have been many proposed methods for obtaining supersymmetric theories via deconstruction in the literature. \cite{Gregoire:2004ic,Nojiri:2004jm,Craig:2014fka}. However, these methods are generally focused on obtaining $D=4$, $\N=2$ supermultiplets from $D=5$ supermultiplets.  As we are not here interested in obtaining BPS representations\footnote{Even supposing that we were, the deconstruction procedure imposed by \ref{sec:sec:3.1} explicitly breaks the deconstructed action's invariance under the BPS operator, $Z = \p_5$; i.e. $\de_Z X = \theta \p_yX$ is no longer a symmetry of the action, which is important because the massive states come from the $D=5$ massless state $(M,0,0,0,M)$, which maintained $Z^2 X = \p_y^2 X = M^2 X$ for BPS state $X$.  The breaking of the y-translation and the need to freeze the vector mode is discussed in \cite{deRham:2014tga}
.}, and instead our interest is in extracting massive $\N=1$ representations, so these methods will not suffice for our purposes.

		\subsubsection{Engineering the Dirac Mass}

We begin by using the simplest fermion, the spin-$\ohalf$ symplectic-Majorana fermion in 5--D:
\ba
	\S &=& \int \d^5x \l( \frac{i}{2} \bPsi^i \, \Ga^{M}\p_M \Psi^i \r) \, , \label{eqn:5ferm} \\
	&=& \int \d^4x\d y \l( \frac{i}{2} \bPsi^i \, \ga^{\mu}\p_{\mu} \Psi^i + \frac{i}{2} \bPsi^i \, (i\gfive) \p_5 \, \Psi^i\r) \, ,
\ea
to build the desired $D=4$ action. A simple application of the deconstruction procedure will not result in the desired Dirac masses, which were required for the massive $U(1)_R$ symmetry in the Zinoviev theory.  To see this, we apply the fermionic descending relations to (\ref{eqn:5ferm}).  This yields an action of the form
\be
	\S_{\text{Dec}} = \int \d^4x\d y \l( \frac{i}{2} \bpsi^i \, \ga^{\mu}\p_{\mu} \psi^i + \frac{1}{2}\ep^{ij} \bpsi^i \p_5 \psi^j\r) \, .
\ee

It is clear then that we need to modify how the $y$-derivative operates on fermionic objects.  It is crucial that the linear fermion theory obeys the Klein-Gordon dispersion relation when the differential operator is ``squared.'' At the level of the field equations in $D=4$ for state $|F\ket$, this is
\ba
	 \l(i \sla{\p} + \epsilon \p_5 \r) |F \ket &=& 0 \nn \\
	 \imply \l( k^{\mu}k_{\mu} + M^2 \r)  |F \ket &=& 0 \nn \, .
\ea
Thus, if we wish to maintain the massive dispersion relation when the fermionic states descend down one dimension, the only crucial thing that the deconstruction substitution must obey is:
\be
	\ep \p_5 \psi  \to m  \Delta \psi \text{ , such that  } m^2 ( \De)^2 =  m^2 \nn \, , \\
\ee
where now $\De$ is not just a matrix on the site basis, but \textit{additionally carries indices for the fermionic flavor basis}.  Thus, if we make use of the fermion flavor indices, we can avoid the disappearance of our fermionic mass terms.  This approach is very much analogous to the original derivation of the Dirac equation where one allows the differential operators to be deformed by matrices, but keep the desired eigenvalues of the squared operations fixed.  This can be predicted merely from trying to get the correct Dirac term for the Zinoviev theory, however, which we explore next.  Since we want a mass term of the form (\ref{eqn:diracmass}), and knowing that
\be
	\De^{ij} = \ep^{ik}\eta^{kj} \, ,   \nn
\ee
tells us that the obvious matrix to deform our derivative by is $\eta$.  Note that crucially $\eta^2 = I$, as needed.
In other words our proposed deconstruction procedure is:
\ba
&&	\psi^i(x,y) \to \psi^i(x)  \, , \nn \\
	&& \p_y \psi^i(x,y) \to m \eta^{ij}\psi^{j}(x) \, , \label{eqn:fermdeconstruction}
\ea
or, equivalently, in terms of 5-D fermions,
\be
	\p_y \Psi^i(x,y) \to m \De^{ij} \Psi^{j}(x)
\ee
with the same integral rule as bosons, where integrals are converted to sums.  Note, these rules can also be established for the symplectic-Majorana variables, but it is more convenient to place them in the 4--D Majorana variables, so we keep to this formulation.

Once these rules are obeyed, the Lagrangian is
\be
	\S_{\text{Dec}} = \int \d^4x \l( i \ohalf \bpsi^i \gamma^{\mu}\p_{\mu} \psi^i + \ohalf m \De^{ij} \bpsi^i \psi^{j} \r) \, ,
\ee
which is the Lagrangian for pair of massive Majorana fermions with a Dirac mass.

	\subsection{Super-Proca Theory a l\'{a} Deconstruction}
	
As a proof of principle for how our deconstruction procedure works for supersymmetric cases, we begin by showing how the super-Proca theory is obtained.  In this, one can illustrate how the deconstruction procedure which preserves a single $D=4$ supersymmetry that explicitly descends from  a linear combination of the $D=5$ supersymmetries.
	
		\subsubsection{The $D=5$ Lagrangian and Super-Transformation Rules}
		
The supersymmetric Maxwell theory in five dimensions is given by a supermultiplet containing one spin-1 field $A_{\mu}$, a symplectic-Majorana fermion $\Psi^i$, and one scalar $\phi$.  In our conventions, the Lagrangian is given by
\be
	\S = \int \d^5x \left[ -\frac{1}{4} F_{MN} F^{MN} + i \frac{1}{2} \bPsi^i \Ga^M \p_M \Psi^i - \frac{1}{2} (\p_M \phi )^2 \right] \, , 
\ee	
and the supersymmetry transformations with global fermionic parameter $\ep^i$ that leave the action invariant are:
\ba
		&& \de A_M = i \bep^i \Ga_M \Psi^i \nn \, ,  \\
		&& \de \Psi^i = -\frac{1}{2} \Ga^{AB} F_{AB} \ep^i -  \Ga^M \p_M \phi \ep^i  \, , \\
		&& \de \phi = i \bep^i \Psi^i  \,\, .
\ea

One can verify that they obey the super-\poin commutation relations
\ba
	&& \lbrack \de_1 , \de_2 \rbrack A_M = 2i \bep^i_2 \Ga^A \ep^i_1 \p_A A_M - \p_M( 2i \bep^i_2 \Ga^A \ep^i_1 A_A ) \nn \, , \\
	&& \lbrack \de_1 , \de_2 \rbrack \Psi^i = 2i \bep^j_2 \Ga^A \ep^j_1 \p_A \Psi^i + \text{(E.O.M.)} \nn \, , \\
	&& \lbrack \de_1 , \de_2 \rbrack \phi = 2i \bep^i_2 \Ga^A \ep^i_1 \p_A \phi \, .
\ea
We need to recast this into manifestly 4-D variables, so we perform a $(4+1)$-split on the action and supertransformations. Beginning with the spin-1 mode, we decompose it into
\ba
	A_M = \bpmat A_{\mu} \\ \pi \epmat \, \, ,
\ea
and use the fermion descending relations \ref{eqn:descform} in Appendix \ref{app:5dspinor} for the fermionic modes.  After applying the (4+1)-split and the descending relations, this translates the 5-D action into the form
\ba
	\S_{\text{5-D sMaxwell}} &=& \int \d^4x \d y \,\left[ -\frac{1}{4}\F_{\mu\nu}\F^{\mu\nu} - \frac{1}{2} (\p_y A_{\mu} - \p_{\mu} \pi )^2 \nn  \right.\\
	&& \,\, +i\frac{1}{2} \bpsi^i \ga^{\mu} \p_{\mu} \psi^i + \frac{1}{2} \ep^{ij} \bpsi^i \p_y \psi^j  \nn \\
	&& \,\, \left.  -\frac{1}{2} (\p_{\mu}\phi)^2 - \frac{1}{2} (\p_y \phi)^2 \right]
\ea
and, given $\ep^i = P^{ij}\vep^j$, places the supertransformations into the form
\ba
	\de A_{\mu} &=& i \bvep^i \ga_{\mu} \psi^i \nn \, , \\
	\de \psi^i &=& -\frac{1}{2} \ga^{\al \beta} \F_{\al\beta} \vep^i - i\ga^{\al} ( \p_y A_{\al} - \p_{\al}\pi ) \ep^{ij}\vep^j  \nn \, , \\
	&& + \ga^{\mu}\gfive\p_{\mu}\phi \, \ep^{ij} \vep^j - i\gfive \p_y\phi \vep^i  \nn \, ,\\
	\de \phi &=&- i \ep^{ij} \bvep^i \gfive \psi^j \nn \, , \\
	\de \pi &=& \ep^{ij} \bvep^i \psi^j \, .
\ea
Notice, crucially, that $\pi$ must enter into the theory as a pseudo spin-0 field if it were a zero mode (e.g. an axion), but since it is going to enter 1-site deconstruction as a \stu field, we shall see the parity-reversal of the lowest bosonic modes.  Thus it will be $\P\T$-even, not odd, which can be read off in the above supertransformations.
		
		\subsubsection{Deconstructing the Action and Transformations}

We now perform deconstruction by applying the 1-site procedures outlined in the previous two sections.  The integral deformation for 1-site deconstruction is trivial, so the complexity is in the derivatives. The procedure amounts to deforming the derivatives as
\ba
&&	\p_y A_{\mu} = m A_{\mu} \, ,  \nn \\
&&	\p_y \psi^i = m \eta^{ij} \psi^j \, , \nn \\
&&	\p_y \phi = m \phi \, .
\ea
Applied to the action, this leads to
\ba
	\S &=& \int \d^4x \,\left[ -\frac{1}{4}\F_{\mu\nu}\F^{\mu\nu} - \frac{1}{2} m^2 A_{\mu} A^{\mu} + m A^{\mu} \p_{\mu} \pi \nn \right. \\
	&& \,\, +i\frac{1}{2} \bpsi^i \ga^{\mu} \p_{\mu} \psi^i + \frac{1}{2}m  \De^{ij }\bpsi^i \psi^j  \nn \\
	&& \,\, \left. - \frac{1}{2} ( \p_{\mu}\pi)^2 -\frac{1}{2} (\p_{\mu}\phi)^2 - \frac{1}{2}m^2 \phi^2 \right]
\ea
which obeys a $U(1)$ \stu symmetry, it has a Dirac mass term for the fermions, and with a comparison to the $\N=2$ vector multiplet action, the helicity decoupling limit $m\to 0$ obeys manifest $\N=2$ SUSY, which is only possible because of the parity reversal amongst the scalars.

Next, we see that the deconstructed transformation laws are
\ba
	\de A_{\mu} &=& i \bvep^i \ga_{\mu} \psi^i  \, ,\nn \\
	\de \psi^i &=& -\frac{1}{2} \ga^{\al \beta} \F_{\al\beta} \vep^i  - i\ga^{\mu}\p_{\mu}\big( \pi -i \gfive  \phi \big) \ep^{ij} \vep^j \nn \, ,  \\
	&& - m \big(i \ga^{\al} A_{\al} \ep^{ij} + i\gfive \phi \de^{ij} \big) \vep^j \nn  \, , \\
	\de \phi &=& -i \ep^{ij} \bvep^i \gfive \psi^j \nn \, , \\
	\de \pi &=& \ep^{ij} \bvep^i \psi^i \, \, ,  \label{decontrans}
\ea
which also are manifestly invariant under the $U(1)$ \stu symmetry.  Additionally, we see that the transformations split up into a pattern $\de_0 + m \de_1$ on the fields.  Since the mass corrections to the transformations exist only for fermions (since they were the only transformations with derivatives), $\de_1$ acts trivially on everything but the fermions.

Superficially, these transformations appear to indicate an $\N=2$ amount of SUSY, but we know at least one supercharge must have been broken by the procedure. This will become apparent in checking the closure (super-Lie algebra) of the transformations. We now turn to the issue of finding the surviving supercharge.
		
		\subsubsection{Deconstructing the Supercharges}

We can see that the maximal amount of supersymmetry coming from a deconstructed theory must be half of the original higher-dimensional supersymmetry.  This follows from the super-\poin algebra and the fact that we have broken the generator $P_y$:
\ba
	\, [ \bep^i_1 Q^i, \bQ^j \ep^j_2] &=& 2i \bep^i_2 \Ga^A \ep^j_1 \, P_A \, \ep^{ij} \nn \\
	&=& 2i \bep^i_2 \big( \Ga^a P_a +  \underbrace{\Ga^5 P_y}_{=0 \text{, must be imposed}} \big)\ep^j_1 \ep^{ij} \nn \\
	& \Rightarrow & \bep^i_2 \ep^{ik }\eta^{kj} \gfive \ep^j_1 = 0 = \De^{ij} \bvep^{i}_2 \vep^j_1 \\
	& \Rightarrow & \vep^i = \al^i \vep  \text{     and      } \De^{ij}\al^i\al^j = 0 \,\, .
\ea
In the above we have made use of our modified fermionic derivative $P_y \ep^i = -i \partial_y \ep^i \rightarrow -i m \eta^{ij} \ep^j$.

We must find the linear combination of supercharges whose anti-commutation algebra does not generate a term proportional to $P_y$ since this generator is broken by the deconstruction procedure, but it must simultaneously leave the action invariant under the transformation.  This linear combination must be such that $ \De^{ij}\al^i\al^j = 0$.
To connect with earlier notation, we define $\beta^i = \De^{ij}\al^j$ so that $\al^i \beta^i = 0 $ which in two dimensions is equivalent to
\be
	\beta^i = c \ep^{ij} \al^j \, , 
\ee
with $c$ being an unknown coefficient.  This imposes the condition that
\be
	\De^{ij} \al^j = c \ep^{ij}\al^j \, .
\ee
Since we can absorb the overall scale of $\al^i$ in the SUSY parameter $\vep$, we can additionally impose $
	\al^i \al^i = 1$. Together these imply
\be	
c = \pm 1   \, , \quad 	\al^i = \bpmat 1 \\ 0 \epmat \, , \quad \beta^i = \bpmat 0 \\ 1 \epmat \, .
\ee
One can verify that upon making this restriction on the supersymmetry transformation, then the supersymmetry variations (\ref{decontrans}) do indeed leave the action invariant provided we choose $c=+1$.  The invariance can be split up into cases by power counting in $m$:
\be
	\de \S = \underbrace{ \de_0 \S_0 }_{=0} + m \Big( \underbrace{ \de_0 \S_1 + \de_1 \S_0 }_{\propto ( \De^{ij} - \ep^{ij} )\al^j} \Big) + m^2 \Big( \underbrace{ \de_1 \S_1 + \de_0 \S_2}_{\propto ( \De^{ij} - \ep^{ij} )\al^j} \Big) + m^3 \underbrace{\de_1 \S_2}_{=0} \, .
\ee
The case $\de_0 \S_0$ is trivially zero, since the kinetic terms and $\de_0$ are $\N=2$ invariant, and $\de_1 \S_2$ is trivially zero the reason outlined.  In this way, we see that a theory for a massive $\N=1$ supermultiplet will arise from a massless $\N=2$ theory. Since this argument is in essence algebraic, it leaves hope that a nonlinear version of this procedure may exist.

\section{Deconstructing Linearized $D=5$ SUGRA to the Zinoviev Theory}
\label{sec: sec4}

\subsection{Reviewing Linearized $D=5$, $\N=2$ Supergravity}

The minimal supermultiplet containing a spin-2 field in 5-D contains one massless spin-2 field $H_{MN}$, one symplectic-Majorana, spin-$\thalf$ field $\Psi_M\,^i$, and one spin-1 field $A_M$ \cite{Cremmer:1980gs,Chamseddine:1980sp,Gunaydin:1983bi,D'Auria:1981kq}.  The action for linearized $D=5$ supergravity is given by the canonical kinetic terms for each of these fields
\be
	\S = \int \d^5 x \left[ -\frac{1}{2} H_M\,^A \de^{MNR}_{ABC} \p_N \p^B H_R\,^C - i \frac{1}{2} \bPsi_M\,^i \Ga^{MNR} \p_N \Psi_R\,^i -\frac{1}{4} F_{MN}F^{MN} \right] \, . 
\ee
The SUSY transformations that leave the action invariant, with group parameter $\ep^i$ which is a symplectic-Majorana fermionic variable, are
\ba
&&	\de H_M\,^A   =  i\frac{1}{2} \bep^i \l(\Ga_M \Psi^A\,^i + \Ga^A \Psi_M\,^i \r) \nn \, ,  \\
&&	\de \Psi_M\,^i  =  \Ga^{AB} \p_A H_{BM} \ep^i + \frac{1}{2\sqrt{6}} \l( \Ga^{AB}\,_M  - 4 \Ga^A \de_M^B \r) F_{AB} \ep^i  \nn \, , \\
&&	\de A_M  =  -i \sqrt{\thalf} \bep^i \Psi_M\,^i \, . 
\ea

	\subsection{Deconstructing to the $\N=1$ Zinoviev Action}

Deconstructing the bosonic fields is well understood, and a straightforward application of the deconstruction procedure leads to the bosonic actions outlined in (\ref{eqn:zinoviev0}), (\ref{eqn:zinoviev1}), and (\ref{eqn:zinoviev2}). Thus we turn to the deconstruction procedure outlined in the previous section applied to the 5--D Rarita-Schwinger action with symplectic-Majorana fermions in order to obtain the appropriate fermionic portion of the action.  We begin by separating the action into its components, and then apply the descending relations to convert it into manifestly 4--D Majorana form:
\ba
	\S_{\text{RS}} &=& \int \d^5x\, \left[ - i \frac{1}{2} \bPsi_M\,^i \Ga^{MNR} \p_N \Psi_R\,^i  \right] \, ,\\
	&=& \int \d^4x\d y\, \l[ - i \frac{1}{2} \bPsi_{\mu}\,^i \ga^{\mu\nu\rho} \p_{\nu} \Psi_{\rho}\,^i - i \frac{1}{2} \bPsi_{\mu}\,^i \ga^{\mu\nu}(i\gfive) \l[2\p_{\nu} \Psi_y\,^i - \p_y \Psi_{\mu}\,^i \r] \r] \, ,  \\
	&=& \int \d^4x\d y\, \l[ - i \frac{1}{2} \bpsi_{\mu}\,^i \ga^{\mu\nu\rho} \p_{\nu} \psi_{\rho}\,^i -\ep^{ij}\bpsi_{\mu}\,^i \ga^{\mu\nu}\p_{\nu} \psi_y\,^i  +\ohalf\ep^{ij }\bpsi_{\mu}\,^i \ga^{\mu\nu} \p_y \psi_{\nu}\,^i \r] \, . \nn \\
\ea
Next, we apply the fermionic deconstruction prescription outlined in (\ref{eqn:fermdeconstruction}), which yields
\be
	\S_{\text{Dex}} = \int \d^4x\, \left[   - i \frac{1}{2} \bpsi_{\mu}\,^i \ga^{\mu\nu\rho} \p_{\nu} \psi_{\rho}\,^i + \frac{1}{2} m \bpsi_{\mu}\,^i  \ga^{\mu\nu}\De^{ij} \psi_{\nu}\,^j - \bpsi_{\mu}\,^i  \ga^{\mu\nu}\ep^{ij} \p_{\nu} \psi_y\,^j \right] \, , 
\ee
and one can cf. to (\ref{eqn:diracmixy}) to see that this is the undiagonalized \stu form of the Rarita-Schwinger-Dirac action, provided that
\be
	\psi_y\,^i = \eta^{ij} \chi^j \, ,
\ee
before rescaling.  Thus we have successfully reproduced the full bosonic and fermionic portions of the $\N=1$ Zinoviev action.  The lingering question is if we are able to see the preserved copy of the supertransformations immediately descending from the broken $\N=2$ supertransformations.

\subsection{Verifying the Zinoviev Super-Transformation Rules}

We immediately use the unmixing transformations (\ref{eqn:helunmixy}) in the definition of the 5-D fields.  This then shows explicitly how the lower dimensional helicity substates of the massive spin fields sit inside of the higher dimensional helicity states.\footnote{A similar kinetic mixing occurs in the slighlty different context for KK compactifications\cite{Chamseddine:1980sp,Gunaydin:1983bi}. Afterwords, all degrees of freedom have canonical kinetic terms. Here, the \stu fields play the role of the lower spin zero-modes, but in massive states they enter as pure gauge modes.}  Then when one deconstructs, one is immediately lead to the canonically normalized massive modes with all helicity states and gauge symmetries made manifest.
We split up the fields into the following 4--D fields, which encode the manifest helicity states of the massive fields:
\ba
&&	H_M\,^A = \bpmat h_{\mu}\,^{\al} - \frac{1}{\rsix} \pi \de_{\mu}^{\al}  & \,\,\, &  \frac{1}{\rtwo}B^{\al} \\ \frac{1}{\rtwo}B_{\mu} & \,\,\, & \sqrt{\frac{2}{3}} \pi   \epmat \, ,  \\
&&	\bar{P}^{ij} \Psi_M\,^j = \bpmat  \psi_{\mu}\,^{i} - \risix \ga_{\mu} \De^{ij}\chi^j \\ \sqrt{\frac{2}{3}}\eta^{ij} \chi^j   \epmat \, , \\
&&	A_M = \bpmat A_{\mu} \\ \ph \epmat \,\, . \label{eqn:4Din5D}
\ea
We will apply this decomposition on the transformations.  First we split the indices into manifestly 4-D Lorentz covariant objects, and then apply the fermionic descending relations to obtain manifestly 4-D spinor objects.  We will begin with the deconstructed spin-2 field's transformations
\ba
	\de H_{y}\,^{y} &=&  -\bep^i \gfive \Psi_y\,^i = \ep^{ij} \bvep^i \l(\rtwothree \eta^{jk}\chi^k\r)\nn  \, , \\
	&=& \rtwothree \de \pi \nn \, , \\
	\imply \de \pi &=& i \beta^i \bvep \chi^i \, , \\
	&& \nn  \, , \\
	\de H_{\mu}\,^y &=& \ohalf \bep^i \l( \ga_{\mu} \Psi_y\,^i + i\gfive \Psi_{\mu}\,^i \r) \nn \, ,  \\
	&=& \frac{1}{\rtwo} \de B_{\mu} \nn \, , \\
	\imply \de B_{\mu} &=& \beta^i \, \frac{1}{\rtwo} \bvep \psi^i + \al^i \,i \frac{\rthree}{2} \bep \ga_{\mu} \chi^i  \, ,  \\
	&& \nn \\
	\de H_{\mu}\,^{\al} &=& i \bep^i \ga_{(\mu}\Psi_{\nu)}\,^i = i \bep^i \ga_{(\mu} \l( \psi_{\nu)}\,^i - \risix \ga_{\nu)} \De^{ij}\chi^j \ \r)  \nn \, ,  \\
	&=& \de h_{\mu}\,^{\al} - \ronesix \de \pi \de_{\mu}^{\al} \nn \, ,  \\
	\imply \de h_{\mu}\,^{\al} &=& \al^i \, i\bvep \ga_{(\mu} \psi_{\nu)}\,^i \, .
\ea
In the final lines, we made use of $\ep^i = \al^i \ep$ and related identities (\ref{eqn:alphaidentities}).  Next we move onto the case for the spin-1 transformations,
\ba
	\de A_y &=& -i \rthreetwo \bep^i \Psi_y\,^i = -i \rthreetwo \ep^{ij} \bvep^i \gfive \l( \rtwothree \eta^{jk} \chi^k \r) \nn \, ,  \\
	\imply \de \ph &=& \beta^i \, i \bvep \gfive \chi^i \,, \\
	&& \\
	\de A_{\mu} &=& -i \rthreetwo \bep^i \Psi_{\mu}\,^i =   -i \rthreetwo \ep^{ij} \bvep^i \gfive \l(  \psi_{\mu}\,^{i} - \risix \ga_{\mu} \De^{ij}\chi^j \r) \nn \, , \\
	\imply \de A_{\mu} &=& \beta^i \, \rthreetwo \bvep \gfive \psi_{\mu}\,^i + \ohalf \bvep \ga_{\mu}\gfive \chi^i \, .
\ea
One may then check that the bosonic transformations are indeed reproduced, with the correct portion of the supercharges broken; cf. (\ref{eqn:zinovievtrans}).

Finally, the fermionic superpartner's transformations must be checked.  This is a more laborious calculation involving repeated use of identities outlined in (\ref{appsec:usefulform}).  We give some the intermediate parts of the calculation here:
\ba
	\bar{P}^{ij} \de\Psi_5\,^j &=& \frac{1}{\rtwo} \ga^{\al\beta} \p_{\al}B_{\beta} - i \ga^{\al}\l( \rtwothree \p_{\al}\pi  - \frac{1}{\rtwo}\p_y B_{\al}\r) \ep^{ij}\vep^j \nn \, , \\
	&& -\frac{1}{2\rsix} \ga^{\al\beta}\gfive \F_{\al\beta}\vep^i + \rtwothree\ga^{\al}\gfive \l( \p_y A_{\al} - \p_{\al}\ph \r) \ep^{ij}\vep^j \nn \, , \\
	&=& \rtwothree \eta^{ij} \de \chi^j \, , 
\ea
and finally the spin-$\thalf$ field:
\ba
	\bar{P}^{ij} \de\Psi_{\mu}\,^{j} &=& \ga^{\al\beta} \p_{\al}\l( h_{\beta\mu} - \ronesix \pi\eta_{\beta\mu} \r) \vep^i + i\ga^{\al}\l[ \p_y\l(h_{\mu\al} - \ronesix \eta^{\mu\al}\r) - \ronetwo B_{\al} \r]\ep^{ij}\vep^j \ \nn \, ,  \\
	&& +\frac{1}{2\rsix} \l[ \ga^{\al\beta}\,_{\mu} -4\ga^{\al}\de^{\beta}_{\mu}\r]\F_{\al\beta} \ep^{ij}\vep^j -\frac{i}{\rsix}[\ga^{\al}\,_{\mu} - 4\de^{\al}_{\mu}]\gfive(\p_yA_{\al}-\p_{\al}\ph)\vep^i \nn \, , \\
	&=& \de \psi_{\mu}\,^i - \risix \ga_{\mu} \De^{ij} \de \chi^j \, . 
\ea
To directly compare these transformations to Zinoviev, we must modify the supersymmetry transformations with compensating supergauge transformations of gauge parameter $(-\ronesix \pi \al^i \vep + i\ronetwo\ga^{\al}B_{\al} \beta^i \vep)$.  Once performed, one can easily read off that the deconstructed super-transformation rules --crucially broken appropriately down the correct $\N=1$ sub-superalgebra-- yields the final, correct super-transformation rules for the fermions:
\ba
	\de \psi_{\mu}\,^i &=& \al^i\, \ga^{\al\beta}\p_{\al} h_{\beta\mu}\vep -\frac{m}{\rtwo} \Big[ \ga_{\mu}\ga^{\al}B_{\al} + i\rthree \gfive A_{\mu} \Big]\al^i \vep \nn \, , \\
	&& -\frac{i}{4\rtwo} \ga^{\al\beta}\ga_{\mu} \Big[ \G_{\al\beta} - \rthree i\gfive \F_{\al\beta}\Big] \beta^i \vep + im \Big[ \ga^{\al}h_{\al\mu} + \ga_{\mu} \pi \Big] \beta^i \vep  \nn \, ,  \\
	&& \nn \\
	\de \chi^i &=& -\frac{1}{4} \ga^{\al\beta} \Big[\rthree \G_{\al\beta} + i\gfive \F_{\al\beta}\Big]\vep \al^i \nn \, ,  \\
	&&    -i \ga^{\al}  \p_{\al} \Big[\pi +\gfive\ph \Big] \beta^i \vep + im \ga^{\al} \Big[ \rthree B_{\al} -i\gfive A_{\al} \Big] \beta^i \vep \, ,
\ea
as promised.  Thus we see that not only is the supersymmetric Lagrangian recovered after deconstruction, but so too can we directly see how and where the supercharge breaks such that a copy of the $\N=1$ massive super-transformations directly descends from the higher dimensional $\N=2$ super-transformations.

\section{Discussion}

By proposing a modified deconstruction procedure for fermionic derivatives, we have shown how at the linear level, a theory of an $\N=1$ massive spin 2 supermuliplet emerges from deconstruction of linearized $\N=2$ supergravity in 5 dimensions. At an algebraic level, this procedure works because when $y$-translations are broken ($P_y$), which is automatically the case in deconstruction, the $N=2$ SUSY algebra in 5 dimensions is broken down to the $\N=1$ super-\poin algebra in 4 dimensions. As an example of this we also show how the 4-D $\N=1$ super-Proca theory arises from 5-D $\N=2$ super-Maxwell theory. This algebraic picture suggests that it may be possible to generalize this to the nonlinear level, and if so it would allow us to give a description of a conjectured interacting theory of a massive spin-2 supermultiplet, in effect a globally supersymmetric extension of massive gravity.  If this does indeed generalize to the non-linear theory, an important question is: does the decoupling limit of the theory uphold the global supersymmetry?  Since the usual massive gravity decoupling limit gives a Galileon theory, this suggests the existence of some super-Galileon theory which necessarily includes spin-1 degrees of freedom, \cite{Ondo:2013wka,Gabadadze:2013ria} (and hence is not necessarily connected with other proposed super-Galileon theories \cite{Khoury:2011da,Farakos:2013zya}).

The massless limit, taken after appropriately canonically normalizing the kinetic terms for all the scalar fields, gives a copy of $\N=2$ linearized supergravity. Just as Lorentz invariant massive gravity breaks the local Diffeomorphism symmetry of GR down to the global \poin group, the conjectured supersymmetric theory will break local $\N=2$ SUSY down to a global $\N=1$ subgroup. This halving of the global supersymmetry is necessary to account for the larger size of the massive supermultiplet. The question of whether such a nonlinear theory exists will be considered in a future work.

What about a locally supersymmetric theory? Since local supersymmetry requires a corresponding massless spin-$\thalf$ gauge field, this in turn necessitates an additional spin-2 field for its superpartner, and thus the fully locally supersymmetric theory must contain two massive spin-2 fields, i.e. it must be a supersymmetric version of a Bi-gravity theory. Since the deconstruction procedure may be easily generalized to give multi-gravity theories, this suggests this may be the most fruitful way to approach the conjectured supersymmetric theories. We leave this to future work.

It is natural to ask where this procedure generalize to higher supersymmetries. The maximal supersymmetry in a massive supergravity theory, subject to the largest state being spin-2, means that after the breaking, the theory will have $\N=4$ supersymmetry in $D=4$.  The linearized limit, the $\N=4$ Zinoviev theory \cite{Zinoviev:2002xn}, is already known as a generalization of the $\N=1$ case. Therefore, one suspects that if one wanted to generate massive supergravity theories with more supersymmetries, one should start by deconstructing $\N=1$ $D=11$ supergravity into a ``master'' $\N=(1,1)$ $D=10$ theory of massive supergravity and then dimensionally reducing to 4 dimensions. We will leave such conjectures for future work.

\acknowledgments

We would like to thank Claudia de Rham and Andrew Matas for discussions. The work of AJT at CWRU was supported by a Department of Energy Early Career Award DE-SC0010600. The work of AJT at IC is supported by STFC grant ST/P000762/1. AJT thanks the Royal Society for support at ICL through a Wolfson Research Merit Award.

\appendix
\section{Conventions and Notations}
\label{sec:appA}

Our conventions are:

\begin{itemize}
\item[1.)] The same as Srednicki \cite{Srednicki:2007qs} (of specific interest are sections 33 through 43).
\item[2.)] Therefore, we use the (-- + ... +) signature, but our gamma matrices obey $\{ \Ga^A, \Ga^A \} = -2 \eta^{AB}$. Grassman numbers obey $(ab)\herm = b\herm a\herm$.
\item[3.)] Our Majorana fermions can easily be converted to Weyl fermions by following the recipe outlined in the above sections of Srednicki.
\item[4.)] All 5-D fermions are given by capital Greek characters (e.g. $\La$, $\Psi$), whereas all 4-D fermions are given by lower case characters (e.g. $\la$, $\psi$).  This will also be true of our bosonic variables, the only exceptions being the vector fields $A_{\mu}$ and $B_{\mu}$.
\item[5.)] Gamma matrices in each dimension are also (un)capitalized respectively, and obey $\Ga^a = \ga^a$ and $\Ga^5 = i \ga_5$. Note that $(\ga_5)^2 = +1$.
\item[6.)] $\Ga^{A_1 \cdots A_n} \eq \frac{1}{n!} \l( \Ga^{A_1} \cdots \Ga^{A_n} + (\text{Perms}) \r)$.
\item[7.)] We make extensive use of the generalized Kronecker delta tensors, e.g. $\de^{A_1 \cdots A_D}_{B_1 \cdots B_D} \eq \ep^{A_1 \cdots A_D} \ep_{B_1 \cdots B_D}$ and $\de^{AB}_{MN} = \de^{A}_{M}\de^{B}_{N} -\de^{A}_{N}\de^{B}_{M}$.  Note that we always have weight 1 (anti)-symmetrization, e.g. $\Ga^{ABC} = \Ga^{[A} \Ga^B \Ga^{C]} = \frac{1}{3!} \de^{ABC}_{MNR} \Ga^M \Ga^M \Ga^R$
\end{itemize}
 We also note that in contrast to many supersymmetry sources in $4$ and $5$ dimensions, our $\bpsi$ will always refer to the Dirac conjugate and never the Majorana conjugate, and the placement of symplectic indices $i$ do \textbf{not} indicate chirality in 4-D.  As such, the height of the symplectic index does not signify anything, and as a convention will always be written upstairs to unclutter notation.

For clarity, we assemble a list of useful $4$x$4$ spinor matrices:
\ba
	\ga^0 &=& \bpmat 0 & I \\ I & 0 \epmat \, , \, \,	\ga^i = \bpmat 0 & \si^i \\ -\si^i & 0 \epmat  \, , \\
	\gfive &=& i \ga^0 \ga^1 \ga^2 \ga^3 = \bpmat -I & 0 \\ 0 & I \epmat \, , \\
	\C_4 &=& -i\ga^0 \ga^2 = \bpmat -\ep & 0 \\ 0 & \ep \epmat \, , \\
	\C_5 &=& \C_4 \gfive = \bpmat \ep & 0 \\ 0 & \ep \epmat \, , \\
	L &=& \frac{1}{2}\l( 1 - \gfive \r) = \bpmat I & 0 \\ 0 & 0 \epmat \, ,  \\ 
	R &=&  \frac{1}{2}\l( 1 + \gfive \r) = \bpmat 0 & 0 \\ 0 & I \epmat \, ,
\ea
and additionally there are useful 2x2 matrices
\ba
	I &=& \de = \bpmat 1 & 0 \\ 0 & 1 \epmat \, , \, \, \si^1 = \De = \bpmat 0 & 1 \\ 1 & 0 \epmat \, , \\
	\si^2 &=& i \ep = \bpmat 0 & -i \\ i & 0 \epmat \, , \, \, \si^3 = \eta = \bpmat 1 & 0 \\ 0 & -1 \epmat \,,  \\
	\ep &=& -i \si^3 = \bpmat 0 & -1 \\ 1 & 0 \epmat \, .
\ea
Note that we will need 2 kinds of two-by-two Hermitian matrices for both the Weyl spinor basis (i.e. $(\si^i)^{\al}\,_{\beta} \, , \, I^{\al}\,_{\beta}$) and the symplectic basis (i.e. $\de^{ij}, \, \eta^{ij}, \De^{ij}$).  We choose to use two different symbols to indicate which basis the matrices are operating on, even though they are equivalent as numerical matrices.

\section{A Review of $5$-D Symplectic-Majorana Spinors}
\label{app:5dspinor}

Although some useful sources exist on the matter, 5-D fermions are generally less well known than their 2, 4, 6, and 10 dimensional counterparts.  Therefore, we pause for a moment to list our conventions for 5-D fermions.

In 4-D, as is well-known, there are 3 distinct kinds of fermions: Weyl fermions, Dirac fermions, and Majorana fermions.  The irreducible representation is Weyl, thus the other two can always be recast as Weyl fermions.  Majorana spinors are Dirac spinors subject to the Majorana constraint:
\ba
	\bpsi &=& \psi^T \C_4 \nn \, ,   \\
  \imply \psi\herm \ga^0 &=& \psi^T \C_4 \, ,  \label{eqn:Maj}
\ea
where $C_4$ is the 4-D charge conjugation matrix.  This kills half of the degrees of freedom of the Dirac fermion, making it a `real' spinor.

By contrast, in 5-D, the only kind of fermions allowed are Dirac.  This means it has 4 degrees of freedom and has no nice Majorana properties to aid calculations.  For this reason, it has become popular in 5-D (and 6-D) supersymmetric theories to make use of an equivalent fermion structure, called a symplectic-Majorana fermion.\footnote{The details are described in \cite{Tanii:1998px,Freedman:2012zz}. The appearance of an $\ep^{ij}$ is due to the fact that in 5-D, the failure to have a Majorana fermion is because without it the conjugate is not a star operator, $\Bar{\bar{\Psi}}^M  = -\Psi$, not the crucial $\Bar{\Bar{\Psi}}^M = \Psi$.  The $\ep$ fixes this problem, but requires the fermion to have an index.}  These fermions are created by taking two Dirac fermions, labeled with an index $i$, and applying the following condition (in analogy to (\ref{eqn:Maj})) to reduce the information back to a single Dirac fermion.  Thus this map must operate on both the symplectic and the spinor basis, which we express as
\be
	\bPsi^i \eq (\Psi^i)\herm \Ga^0 = (\Psi^j)^T \ep^{ji} \C_5  \label{eqn:spMaj} \, , 
\ee
where $\C_5$ is the 5-D charge conjugation matrix.

Since in 4-D, we are interested in Majorana fermions and in 5-D we are interested in symplectic-Majorana fermions, an important question is how the 4-D Majorana fermions sit inside their higher dimensional representations. We therefore construct a map between the two 4-D states, which are labeled by a symplectic index, and the 5-D states:
\be
		\Psi^i = P^{ij}\psi^j \, .
\ee
Plugging this relation into (\ref{eqn:spMaj}) and using (\ref{eqn:Maj}), we see that one solution (and a useful one) is to set:
\be
	P^{ij} = \frac{1}{\rtwo}\Big[I\de^{ij} - \gfive \ep^{ij} \Big] \label{eqn:desc} \, , 
\ee
recalling that $I$ and $\gfive$ operate on the spinor basis, but $\de$ and $\ep$ operate on the symplectic indices.  Then the inverse is given by
\be
	\psi^i = \bar{P}^{ij} \Psi^j = \frac{1}{\rtwo}\Big[ I \de^{ij} + \gfive \ep^{ij} \Big] \Psi^j \, .
\ee
The following formulas can be verified from the previous matrices, and are useful:
\ba
\bPsi^i &=& \bpsi^j P^{ji}  \, , \\
\bPsi^i N \Psi^i &=& \begin{cases} \bpsi^i N \psi^i &\mbox{ if } \{ N , \, \gfive \} =0 \, , \\ 
- \ep^{ij} \bpsi^i N \gfive \psi^j &\mbox{ if } [ N , \, \gfive ]=0 \, ,  \end{cases}  \\
P^{ij}\Psi^j &=& -\gfive \ep^{ij}\psi^j \label{eqn:descform} \, , \\
\bar{P}^{ij} \Psi^j &=& \psi^i \, , 
\ea
where $N$ is taken to be an arbitrary spinor matrix.  Note that bispinor products are $\C\P$-odd, because they will couple to $\C\P$-odd fields after deconstruction.

\section{Useful Formulas}
\label{appsec:usefulform}

Here we assemble some useful bosonic relations, recalling that $X^M = (x^{\mu}, y)$. We make copious use of the generalized Kronecker delta tensors, which are the tensor formulation of weighted, anti-syummetric permutations of indices, i.e.
\be
	T_{[\mu_1 \cdots \mu_n]} = \frac{1}{n!} \de^{\mu'_1 \cdots \mu'_n}_{\mu_1 \cdots \mu_n} T_{\mu'_1 \cdots \mu'_n} \,\, .
\ee
We begin with needed recursion relations for the generalized Kronecker delta (they are dimension independent, these hold true for D-dim generalized Kronecker deltas):
\ba
		\de^{\mu\nu\rho}_{\al\beta\ga} &=& \de^{\mu}_{\al} \de^{\nu\rho}_{\beta \ga} + \de^{\mu}_{\beta} \de^{\nu\rho}_{\ga \al} + \de^{\mu}_{\ga} \de^{\nu\rho}_{\al \beta} \, ,  \\
		\de^{\mu\nu}_{\al\beta} &=& \de^{\mu}_{\al} \de^{\nu}_{\beta} + \de^{\mu}_{\beta} \de^{\nu}_{\al} \, .
\ea
The contraction identities in dimension $D$, i.e. $\de^{A_1 \cdots A_{d-n} C_1 \cdots C_k}_{B_1 \cdots B_{d-n} C_1 \cdots C_k} = k! \, \de^{A_1 \cdots A_{d-n}}_{B_1 \cdots B_{d-n}}$, follow simple a simple pattern:
\ba
\begin{matrix}
	\text{ 5-D }&\,\,\,\,\,\,\,\,\,&& \text{ 4-D} \nn \\
	\de^{MNRS \bullet}_{ABCD \bullet } = (1)\de^{MNRS}_{ABCD} &&& \de^{\mu\nu\rho \bullet}_{\al\beta\ga\bullet} = (1)\de^{\mu\nu\rho}_{\al\beta\ga} \nn \\
	\nn \\
	\de^{MNR \bullet}_{ABC \bullet } = (2)\de^{MNR}_{ABC} &&& \de^{\mu\nu\bullet}_{\al\beta\bullet} = (2)\de^{\mu\nu}_{\al\beta} \nn \\
	\vdots &&&  \nn \\
	\de^{M \bullet}_{A \bullet } = (4)\de^{M}_{A} &&& \de^{\mu\bullet}_{\al\bullet} = (3)\de^{\mu}_{\al} \nn \\ 
	\nn \\
	\de^{M}_{M} = 5 &&& \de^{\mu}_{\mu} = 4
\end{matrix}
\ea
Next we write down some useful gamma matrix relations, where again these relations hold true in any dimension.  We begin with a few simple recursion relations:
\ba
	\Ga^{ABC} &=& \frac{1}{3!} \de^{ABC}_{MNR} \Ga^M \Ga^M \Ga^R \nn\\
	&=& \frac{1}{3} \l(\Ga^{A}\Ga^{BC} + \Ga^{B}\Ga^{CA} + \Ga^{C}\Ga^{AB}\r) \nn \\
	&=& \Ga^A\Ga^B\Ga^C + \Ga^A \eta^{BC} - \Ga^B \eta^{CA} + \Ga^C \eta^{AB} \, ,   \\
	\nn  \\
	\Ga^{AB} &=& \frac{1}{2!} \l( \Ga^A \Ga^B - \Ga^B \Ga^A \r) = \Ga^A\Ga^B + \eta^{AB} \, .
\ea
Then the dimension-dependent contraction identities are given by
\ba
\begin{matrix}
	&\text{ 5-D }& &\,\,\,\,\,\,\,\,\,&&  &\text{ 4-D}& \nn \\
	\Ga^{ABC}\Ga_C &=& -3\Ga^{AB}&&& \ga^{abc}\ga_c &=& -2\ga^{ab} \\
	\Ga^{AB}\Ga_B &=& -4\Ga^A &&& \ga^{ab}\ga_b &=& -3\ga^a  \nn \\
	\Ga^A \Ga_A &=& -5 &&& \ga^a \ga_a &=& -4
\end{matrix}
\ea
Noting that this holds true for contractions on either side (i.e. $\ga^{a\bullet}\ga_{\bullet} = \ga_{\bullet} \ga^{\bullet a}$).
Some useful identities of the flavor matrices follow from them being a representation of $\spin (3)$:
\ba
	\De^{ij} \eta^{jk} &=& \ep^{ik}= -\eta^{ij} \De^{jk}  \, ,  \\
	\ep^{ij}\eta^{jk} &=& \De^{ik} = -\eta^{ij}\ep^{jk} \, , \\
	\De^{ij} \ep^{jk} &=& \eta^{ik}=  -\ep^{ij}\De^{jk} \, .
\ea

\bibliographystyle{JHEPmodplain}
\bibliography{refs}

\providecommand{\href}[2]{#2}\begingroup\raggedright\begin{thebibliography}{10}

\bibitem{deRham:2010gu}
C.~de~Rham and G.~Gabadadze, {\it {Selftuned Massive Spin-2}},  {\sl
  Phys.Lett.} {\bf B693} (2010) 334--338,
  [\href{http://arxiv.org/abs/1006.4367}{{\sf arXiv:1006.4367}}],
  [\href{http://dx.doi.org/10.1016/j.physletb.2010.08.043}{{\sf
  doi:10.1016/j.physletb.2010.08.043}}].

\bibitem{deRham:2010ik}
C.~de~Rham and G.~Gabadadze, {\it {Generalization of the Fierz-Pauli Action}},
  {\sl Phys.Rev.} {\bf D82} (2010) 044020,
  [\href{http://arxiv.org/abs/1007.0443}{{\sf arXiv:1007.0443}}],
  [\href{http://dx.doi.org/10.1103/PhysRevD.82.044020}{{\sf
  doi:10.1103/PhysRevD.82.044020}}].

\bibitem{deRham:2010kj}
C.~de~Rham, G.~Gabadadze, and A.~J. Tolley, {\it {Resummation of Massive
  Gravity}},  {\sl Phys.Rev.Lett.} {\bf 106} (2011) 231101,
  [\href{http://arxiv.org/abs/1011.1232}{{\sf arXiv:1011.1232}}],
  [\href{http://dx.doi.org/10.1103/PhysRevLett.106.231101}{{\sf
  doi:10.1103/PhysRevLett.106.231101}}].

\bibitem{Hassan:2011hr}
S.~F. Hassan and R.~A. Rosen, {\it {Resolving the Ghost Problem in non-Linear
  Massive Gravity}},  {\sl Phys. Rev. Lett.} {\bf 108} (2012) 041101,
  [\href{http://arxiv.org/abs/1106.3344}{{\sf arXiv:1106.3344}}],
  [\href{http://dx.doi.org/10.1103/PhysRevLett.108.041101}{{\sf
  doi:10.1103/PhysRevLett.108.041101}}].

\bibitem{deRham:2014zqa}
C.~de~Rham, {\it {Massive Gravity}},
  \href{http://arxiv.org/abs/1401.4173}{{\sf arXiv:1401.4173}}.

\bibitem{Nicolis:2008in}
A.~Nicolis, R.~Rattazzi, and E.~Trincherini, {\it {The Galileon as a local
  modification of gravity}},  {\sl Phys.Rev.} {\bf D79} (2009) 064036,
  [\href{http://arxiv.org/abs/0811.2197}{{\sf arXiv:0811.2197}}],
  [\href{http://dx.doi.org/10.1103/PhysRevD.79.064036}{{\sf
  doi:10.1103/PhysRevD.79.064036}}].

\bibitem{deRham:2012ew}
C.~de~Rham, G.~Gabadadze, L.~Heisenberg, and D.~Pirtskhalava, {\it
  {Non-Renormalization and Naturalness in a Class of Scalar-Tensor Theories}},
  {\sl Phys.Rev.} {\bf D87} (2013) 085017,
  [\href{http://arxiv.org/abs/1212.4128}{{\sf arXiv:1212.4128}}],
  [\href{http://dx.doi.org/10.1103/PhysRevD.87.085017}{{\sf
  doi:10.1103/PhysRevD.87.085017}}].

\bibitem{deRham:2013qqa}
C.~de~Rham, L.~Heisenberg, and R.~H. Ribeiro, {\it {Quantum Corrections in
  Massive Gravity}},  {\sl Phys.Rev.} {\bf D88} (2013) 084058,
  [\href{http://arxiv.org/abs/1307.7169}{{\sf arXiv:1307.7169}}],
  [\href{http://dx.doi.org/10.1103/PhysRevD.88.084058}{{\sf
  doi:10.1103/PhysRevD.88.084058}}].

\bibitem{Cheung:2016yqr}
C.~Cheung and G.~N. Remmen, {\it {Positive Signs in Massive Gravity}},  {\sl
  JHEP} {\bf 04} (2016) 002, [\href{http://arxiv.org/abs/1601.04068}{{\sf
  arXiv:1601.04068}}], [\href{http://dx.doi.org/10.1007/JHEP04(2016)002}{{\sf
  doi:10.1007/JHEP04(2016)002}}].

\bibitem{Keltner:2015xda}
L.~Keltner and A.~J. Tolley, {\it {UV properties of Galileons: Spectral
  Densities}},  \href{http://arxiv.org/abs/1502.05706}{{\sf arXiv:1502.05706}}.

\bibitem{Malaeb:2013lia}
O.~Malaeb, {\it {Massive Gravity with $N=1$ local Supersymmetry}},  {\sl Eur.
  Phys. J.} {\bf C73} (2013), no.~9 2549,
  [\href{http://arxiv.org/abs/1302.5092}{{\sf arXiv:1302.5092}}],
  [\href{http://dx.doi.org/10.1140/epjc/s10052-013-2549-9}{{\sf
  doi:10.1140/epjc/s10052-013-2549-9}}].

\bibitem{Malaeb:2013nra}
O.~Malaeb, {\it {Supersymmetrizing Massive Gravity}},  {\sl Phys. Rev.} {\bf
  D88} (2013), no.~2 025002, [\href{http://arxiv.org/abs/1303.3580}{{\sf
  arXiv:1303.3580}}], [\href{http://dx.doi.org/10.1103/PhysRevD.88.025002}{{\sf
  doi:10.1103/PhysRevD.88.025002}}].

\bibitem{Chamseddine:2010ub}
A.~H. Chamseddine and V.~Mukhanov, {\it {Higgs for Graviton: Simple and Elegant
  Solution}},  {\sl JHEP} {\bf 08} (2010) 011,
  [\href{http://arxiv.org/abs/1002.3877}{{\sf arXiv:1002.3877}}],
  [\href{http://dx.doi.org/10.1007/JHEP08(2010)011}{{\sf
  doi:10.1007/JHEP08(2010)011}}].

\bibitem{DelMonte:2016czb}
F.~Del~Monte, D.~Francia, and P.~A. Grassi, {\it {Multimetric Supergravities}},
   {\sl JHEP} {\bf 09} (2016) 064, [\href{http://arxiv.org/abs/1605.06793}{{\sf
  arXiv:1605.06793}}], [\href{http://dx.doi.org/10.1007/JHEP09(2016)064}{{\sf
  doi:10.1007/JHEP09(2016)064}}].

\bibitem{Zinoviev:2018juc}
Y.~M. Zinoviev, {\it {On massive super(bi)gravity in the constructive
  approach}},  \href{http://arxiv.org/abs/1805.01650}{{\sf arXiv:1805.01650}}.

\bibitem{Garcia-Saenz:2018wnw}
S.~Garcia-Saenz, K.~Hinterbichler, and R.~A. Rosen, {\it {Supersymmetric
  Partially Massless Fields and Non-Unitary Superconformal Representations}},
  \href{http://arxiv.org/abs/1810.01881}{{\sf arXiv:1810.01881}}.

\bibitem{Fierz:1939ix}
M.~Fierz and W.~Pauli, {\it {On relativistic wave equations for particles of
  arbitrary spin in an electromagnetic field}},  {\sl Proc.Roy.Soc.Lond.} {\bf
  A173} (1939) 211--232, [\href{http://dx.doi.org/10.1098/rspa.1939.0140}{{\sf
  doi:10.1098/rspa.1939.0140}}].

\bibitem{Fierz:1939zz}
M.~Fierz, {\it {Force-free particles with any spin}},  {\sl Helv.Phys.Acta}
  {\bf 12} (1939) 3--37.

\bibitem{vanDam:1970vg}
H.~van Dam and M.~Veltman, {\it {Massive and massless Yang-Mills and
  gravitational fields}},  {\sl Nucl.Phys.} {\bf B22} (1970) 397--411,
  [\href{http://dx.doi.org/10.1016/0550-3213(70)90416-5}{{\sf
  doi:10.1016/0550-3213(70)90416-5}}].

\bibitem{Zakharov:1970cc}
V.~Zakharov, {\it {Linearized gravitation theory and the graviton mass}},  {\sl
  JETP Lett.} {\bf 12} (1970) 312.

\bibitem{Vainshtein:1972sx}
A.~Vainshtein, {\it {To the problem of nonvanishing gravitation mass}},  {\sl
  Phys.Lett.} {\bf B39} (1972) 393--394,
  [\href{http://dx.doi.org/10.1016/0370-2693(72)90147-5}{{\sf
  doi:10.1016/0370-2693(72)90147-5}}].

\bibitem{Babichev:2013usa}
E.~Babichev and C.~Deffayet, {\it {An introduction to the Vainshtein
  mechanism}},  {\sl Class.Quant.Grav.} {\bf 30} (2013) 184001,
  [\href{http://arxiv.org/abs/1304.7240}{{\sf arXiv:1304.7240}}],
  [\href{http://dx.doi.org/10.1088/0264-9381/30/18/184001}{{\sf
  doi:10.1088/0264-9381/30/18/184001}}].

\bibitem{Boulware:1973my}
D.~Boulware and S.~Deser, {\it {Can gravitation have a finite range?}},  {\sl
  Phys.Rev.} {\bf D6} (1972) 3368--3382,
  [\href{http://dx.doi.org/10.1103/PhysRevD.6.3368}{{\sf
  doi:10.1103/PhysRevD.6.3368}}].

\bibitem{Creminelli:2005qk}
P.~Creminelli, A.~Nicolis, M.~Papucci, and E.~Trincherini, {\it {Ghosts in
  massive gravity}},  {\sl JHEP} {\bf 0509} (2005) 003,
  [\href{http://arxiv.org/abs/hep-th/0505147}{{\sf arXiv:hep-th/0505147}}],
  [\href{http://dx.doi.org/10.1088/1126-6708/2005/09/003}{{\sf
  doi:10.1088/1126-6708/2005/09/003}}].

\bibitem{Aragone:1971kh}
C.~Aragone and S.~Deser, {\it {Constraints on gravitationally coupled tensor
  fields}},  {\sl Nuovo Cim.} {\bf A3} (1971) 709--720,
  [\href{http://dx.doi.org/10.1007/BF02813572}{{\sf doi:10.1007/BF02813572}}].

\bibitem{Aragone:1979bm}
C.~Aragone and S.~Deser, {\it {Consistency Problems of Spin-2 Gravity
  Coupling}},  {\sl Nuovo Cim.} {\bf B57} (1980) 33--49,
  [\href{http://dx.doi.org/10.1007/BF02722400}{{\sf doi:10.1007/BF02722400}}].

\bibitem{Hassan:2011ea}
S.~Hassan and R.~A. Rosen, {\it {Confirmation of the Secondary Constraint and
  Absence of Ghost in Massive Gravity and Bimetric Gravity}},  {\sl JHEP} {\bf
  1204} (2012) 123, [\href{http://arxiv.org/abs/1111.2070}{{\sf
  arXiv:1111.2070}}], [\href{http://dx.doi.org/10.1007/JHEP04(2012)123}{{\sf
  doi:10.1007/JHEP04(2012)123}}].

\bibitem{Hassan:2011tf}
S.~Hassan, R.~A. Rosen, and A.~Schmidt-May, {\it {Ghost-free Massive Gravity
  with a General Reference Metric}},  {\sl JHEP} {\bf 1202} (2012) 026,
  [\href{http://arxiv.org/abs/1109.3230}{{\sf arXiv:1109.3230}}],
  [\href{http://dx.doi.org/10.1007/JHEP02(2012)026}{{\sf
  doi:10.1007/JHEP02(2012)026}}].

\bibitem{Mirbabayi:2011aa}
M.~Mirbabayi, {\it {A Proof Of Ghost Freedom In de Rham-Gabadadze-Tolley
  Massive Gravity}},  {\sl Phys.Rev.} {\bf D86} (2012) 084006,
  [\href{http://arxiv.org/abs/1112.1435}{{\sf arXiv:1112.1435}}],
  [\href{http://dx.doi.org/10.1103/PhysRevD.86.084006}{{\sf
  doi:10.1103/PhysRevD.86.084006}}].

\bibitem{ArkaniHamed:2001nc}
N.~Arkani-Hamed, A.~G. Cohen, and H.~Georgi, {\it {Electroweak symmetry
  breaking from dimensional deconstruction}},  {\sl Phys.Lett.} {\bf B513}
  (2001) 232--240, [\href{http://arxiv.org/abs/hep-ph/0105239}{{\sf
  arXiv:hep-ph/0105239}}],
  [\href{http://dx.doi.org/10.1016/S0370-2693(01)00741-9}{{\sf
  doi:10.1016/S0370-2693(01)00741-9}}].

\bibitem{ArkaniHamed:2002sp}
N.~Arkani-Hamed, H.~Georgi, and M.~D. Schwartz, {\it {Effective field theory
  for massive gravitons and gravity in theory space}},  {\sl Annals Phys.} {\bf
  305} (2003) 96--118, [\href{http://arxiv.org/abs/hep-th/0210184}{{\sf
  arXiv:hep-th/0210184}}],
  [\href{http://dx.doi.org/10.1016/S0003-4916(03)00068-X}{{\sf
  doi:10.1016/S0003-4916(03)00068-X}}].

\bibitem{Schwartz:2003vj}
M.~D. Schwartz, {\it {Constructing gravitational dimensions}},  {\sl Phys.Rev.}
  {\bf D68} (2003) 024029, [\href{http://arxiv.org/abs/hep-th/0303114}{{\sf
  arXiv:hep-th/0303114}}],
  [\href{http://dx.doi.org/10.1103/PhysRevD.68.024029}{{\sf
  doi:10.1103/PhysRevD.68.024029}}].

\bibitem{ArkaniHamed:2003vb}
N.~Arkani-Hamed and M.~D. Schwartz, {\it {Discrete gravitational dimensions}},
  {\sl Phys.Rev.} {\bf D69} (2004) 104001,
  [\href{http://arxiv.org/abs/hep-th/0302110}{{\sf arXiv:hep-th/0302110}}],
  [\href{http://dx.doi.org/10.1103/PhysRevD.69.104001}{{\sf
  doi:10.1103/PhysRevD.69.104001}}].

\bibitem{Chamseddine:1980sp}
A.~H. Chamseddine and H.~Nicolai, {\it {Coupling the SO(2) Supergravity Through
  Dimensional Reduction}},  {\sl Phys. Lett.} {\bf B96} (1980) 89,
  [\href{http://dx.doi.org/10.1016/0370-2693(80)90218-X}{{\sf
  doi:10.1016/0370-2693(80)90218-X}}].

\bibitem{Dolan:1983aa}
L.~Dolan and M.~J. Duff, {\it {Kac-moody Symmetries of {Kaluza-Klein}
  Theories}},  {\sl Phys. Rev. Lett.} {\bf 52} (1984) 14,
  [\href{http://dx.doi.org/10.1103/PhysRevLett.52.14}{{\sf
  doi:10.1103/PhysRevLett.52.14}}].

\bibitem{Dolan:1984fm}
L.~Dolan, {\it {Symmetries of Massive Fields in {Kaluza-Klein} Supergravity}},
  {\sl Phys. Rev.} {\bf D30} (1984) 2474,
  [\href{http://dx.doi.org/10.1103/PhysRevD.30.2474}{{\sf
  doi:10.1103/PhysRevD.30.2474}}].

\bibitem{Deffayet:2003zk}
C.~Deffayet and J.~Mourad, {\it {Multigravity from a discrete extra
  dimension}},  {\sl Phys.Lett.} {\bf B589} (2004) 48--58,
  [\href{http://arxiv.org/abs/hep-th/0311124}{{\sf arXiv:hep-th/0311124}}],
  [\href{http://dx.doi.org/10.1016/j.physletb.2004.03.053}{{\sf
  doi:10.1016/j.physletb.2004.03.053}}].

\bibitem{Deffayet:2004ws}
C.~Deffayet and J.~Mourad, {\it {Some properties of multigravity theories and
  discretized brane worlds}},  {\sl Int.J.Theor.Phys.} {\bf 43} (2004)
  855--864, [\href{http://dx.doi.org/10.1023/B:IJTP.0000048176.15115.f3}{{\sf
  doi:10.1023/B:IJTP.0000048176.15115.f3}}].

\bibitem{Deffayet:2005yn}
C.~Deffayet and J.~Mourad, {\it {Deconstruction of gravity}},  {\sl
  Int.J.Theor.Phys.} {\bf 44} (2005) 1743--1752,
  [\href{http://dx.doi.org/10.1007/s10773-005-8892-0}{{\sf
  doi:10.1007/s10773-005-8892-0}}].

\bibitem{deRham:2013awa}
C.~de~Rham, A.~Matas, and A.~J. Tolley, {\it {Deconstructing Dimensions and
  Massive Gravity}},  \href{http://arxiv.org/abs/1308.4136}{{\sf
  arXiv:1308.4136}}.

\bibitem{Hinterbichler:2012cn}
K.~Hinterbichler and R.~A. Rosen, {\it {Interacting Spin-2 Fields}},  {\sl
  JHEP} {\bf 1207} (2012) 047, [\href{http://arxiv.org/abs/1203.5783}{{\sf
  arXiv:1203.5783}}], [\href{http://dx.doi.org/10.1007/JHEP07(2012)047}{{\sf
  doi:10.1007/JHEP07(2012)047}}].

\bibitem{Deffayet:2012zc}
C.~Deffayet, J.~Mourad, and G.~Zahariade, {\it {A note on 'symmetric' vielbeins
  in bimetric, massive, perturbative and non perturbative gravities}},  {\sl
  JHEP} {\bf 1303} (2013) 086, [\href{http://arxiv.org/abs/1208.4493}{{\sf
  arXiv:1208.4493}}], [\href{http://dx.doi.org/10.1007/JHEP03(2013)086}{{\sf
  doi:10.1007/JHEP03(2013)086}}].

\bibitem{Deser:1974cy}
S.~Deser and P.~van Nieuwenhuizen, {\it {Nonrenormalizability of the Quantized
  Dirac-Einstein System}},  {\sl Phys.Rev.} {\bf D10} (1974) 411,
  [\href{http://dx.doi.org/10.1103/PhysRevD.10.411}{{\sf
  doi:10.1103/PhysRevD.10.411}}].

\bibitem{Ondo:2013wka}
N.~A. Ondo and A.~J. Tolley, {\it {Complete Decoupling Limit of Ghost-free
  Massive Gravity}},  {\sl JHEP} {\bf 1311} (2013) 059,
  [\href{http://arxiv.org/abs/1307.4769}{{\sf arXiv:1307.4769}}],
  [\href{http://dx.doi.org/10.1007/JHEP11(2013)059}{{\sf
  doi:10.1007/JHEP11(2013)059}}].

\bibitem{Hassan:2012wt}
S.~F. Hassan, A.~Schmidt-May, and M.~von Strauss, {\it {Metric Formulation of
  Ghost-Free Multivielbein Theory}},
  \href{http://arxiv.org/abs/1204.5202}{{\sf arXiv:1204.5202}}.

\bibitem{Quevedo:2010ui}
F.~Quevedo, S.~Krippendorf, and O.~Schlotterer, {\it {Cambridge Lectures on
  Supersymmetry and Extra Dimensions}},
  \href{http://arxiv.org/abs/1011.1491}{{\sf arXiv:1011.1491}}.

\bibitem{Tanii:1998px}
Y.~Tanii, {\it {Introduction to supergravities in diverse dimensions}},  in
  {\em {YITP Workshop on Supersymmetry Kyoto, Japan, March 27-30, 1996}}, 1998.
\newblock \href{http://arxiv.org/abs/hep-th/9802138}{{\sf
  arXiv:hep-th/9802138}}.

\bibitem{deWit:2002vz}
B.~de~Wit, {\it {Supergravity}},  in {\em {Unity from duality: Gravity, gauge
  theory and strings. Proceedings, NATO Advanced Study Institute, Euro Summer
  School, 76th session, Les Houches, France, July 30-August 31, 2001}},
  pp.~1--135, 2002.
\newblock \href{http://arxiv.org/abs/hep-th/0212245}{{\sf
  arXiv:hep-th/0212245}}.

\bibitem{Gupta:1954zz}
S.~N. Gupta, {\it {Gravitation and Electromagnetism}},  {\sl Phys.Rev.} {\bf
  96} (1954) 1683--1685, [\href{http://dx.doi.org/10.1103/PhysRev.96.1683}{{\sf
  doi:10.1103/PhysRev.96.1683}}].

\bibitem{Feynman:1996kb}
R.~Feynman, F.~Morinigo, W.~Wagner, and B.~Hatfield, {\em {Feynman lectures on
  gravitation}}.
\newblock Addison-Wesley, 1996.

\bibitem{Weinberg:1965rz}
S.~Weinberg, {\it {Photons and gravitons in perturbation theory: Derivation of
  Maxwell's and Einstein's equations}},  {\sl Phys.Rev.} {\bf 138} (1965)
  B988--B1002, [\href{http://dx.doi.org/10.1103/PhysRev.138.B988}{{\sf
  doi:10.1103/PhysRev.138.B988}}].

\bibitem{Deser:1969wk}
S.~Deser, {\it {Self--interaction and gauge invariance}},  {\sl Gen.Rel.Grav.}
  {\bf 1} (1970) 9--18, [\href{http://arxiv.org/abs/gr-qc/0411023}{{\sf
  arXiv:gr-qc/0411023}}], [\href{http://dx.doi.org/10.1007/BF00759198}{{\sf
  doi:10.1007/BF00759198}}].

\bibitem{Boulware:1974sr}
D.~G. Boulware and S.~Deser, {\it {Classical General Relativity Derived from
  Quantum Gravity}},  {\sl Annals Phys.} {\bf 89} (1975) 193,
  [\href{http://dx.doi.org/10.1016/0003-4916(75)90302-4}{{\sf
  doi:10.1016/0003-4916(75)90302-4}}].

\bibitem{Kiriushcheva:2009tg}
N.~Kiriushcheva and S.~V. Kuzmin, {\it {The Hamiltonian formulation of N-bein,
  Einstein-Cartan, gravity in any dimension: The Progress Report}},  in {\em
  {CAIMS * SCMAI 2009: 30th anniversary of the Canadian Applied and Industrial
  Mathematics Society London, Ontario, Canada, June 10-14, 2009}}, 2009.
\newblock \href{http://arxiv.org/abs/0907.1553}{{\sf arXiv:0907.1553}}.

\bibitem{Kiriushcheva:2009nj}
N.~Kiriushcheva and S.~V. Kuzmin, {\it {Darboux coordinates for the Hamiltonian
  of first order Einstein-Cartan gravity}},  {\sl Int. J. Theor. Phys.} {\bf
  49} (2010) 2859--2890, [\href{http://arxiv.org/abs/0912.5490}{{\sf
  arXiv:0912.5490}}], [\href{http://dx.doi.org/10.1007/s10773-010-0479-y}{{\sf
  doi:10.1007/s10773-010-0479-y}}].

\bibitem{Freedman:1976xh}
D.~Z. Freedman, P.~van Nieuwenhuizen, and S.~Ferrara, {\it {Progress Toward a
  Theory of Supergravity}},  {\sl Phys. Rev.} {\bf D13} (1976) 3214--3218,
  [\href{http://dx.doi.org/10.1103/PhysRevD.13.3214}{{\sf
  doi:10.1103/PhysRevD.13.3214}}].

\bibitem{Deser:1976eh}
S.~Deser and B.~Zumino, {\it {Consistent Supergravity}},  {\sl Phys. Lett.}
  {\bf B62} (1976) 335,
  [\href{http://dx.doi.org/10.1016/0370-2693(76)90089-7}{{\sf
  doi:10.1016/0370-2693(76)90089-7}}].

\bibitem{Grisaru:1976vm}
M.~T. Grisaru, H.~N. Pendleton, and P.~van Nieuwenhuizen, {\it {Supergravity
  and the S Matrix}},  {\sl Phys. Rev.} {\bf D15} (1977) 996,
  [\href{http://dx.doi.org/10.1103/PhysRevD.15.996}{{\sf
  doi:10.1103/PhysRevD.15.996}}].

\bibitem{Gabadadze:2013ria}
G.~Gabadadze, K.~Hinterbichler, D.~Pirtskhalava, and Y.~Shang, {\it {Potential
  for general relativity and its geometry}},  {\sl Phys. Rev.} {\bf D88}
  (2013), no.~8 084003, [\href{http://arxiv.org/abs/1307.2245}{{\sf
  arXiv:1307.2245}}], [\href{http://dx.doi.org/10.1103/PhysRevD.88.084003}{{\sf
  doi:10.1103/PhysRevD.88.084003}}].

\bibitem{Zinoviev:2002xn}
{\relax Yu}.~M. Zinoviev, {\it {Massive spin two supermultiplets}},
  \href{http://arxiv.org/abs/hep-th/0206209}{{\sf arXiv:hep-th/0206209}}.

\bibitem{Gregoire:2004ic}
T.~Gregoire, M.~D. Schwartz, and Y.~Shadmi, {\it {Massive supergravity and
  deconstruction}},  {\sl JHEP} {\bf 07} (2004) 029,
  [\href{http://arxiv.org/abs/hep-th/0403224}{{\sf arXiv:hep-th/0403224}}],
  [\href{http://dx.doi.org/10.1088/1126-6708/2004/07/029}{{\sf
  doi:10.1088/1126-6708/2004/07/029}}].

\bibitem{Nojiri:2004jm}
S.~Nojiri and S.~D. Odintsov, {\it {Multisupergravity from latticized extra
  dimension}},  {\sl Phys. Lett.} {\bf B590} (2004) 295--302,
  [\href{http://arxiv.org/abs/hep-th/0403162}{{\sf arXiv:hep-th/0403162}}],
  [\href{http://dx.doi.org/10.1016/j.physletb.2004.03.078}{{\sf
  doi:10.1016/j.physletb.2004.03.078}}].

\bibitem{Craig:2014fka}
N.~Craig and H.~K. Lou, {\it {Scherk-Schwarz Supersymmetry Breaking in 4D}},
  {\sl JHEP} {\bf 12} (2014) 184, [\href{http://arxiv.org/abs/1406.4880}{{\sf
  arXiv:1406.4880}}], [\href{http://dx.doi.org/10.1007/JHEP12(2014)184}{{\sf
  doi:10.1007/JHEP12(2014)184}}].

\bibitem{deRham:2014tga}
C.~de~Rham, A.~Matas, N.~Ondo, and A.~J. Tolley, {\it {Interactions of Charged
  Spin-2 Fields}},  {\sl Class. Quant. Grav.} {\bf 32} (2015), no.~17 175008,
  [\href{http://arxiv.org/abs/1410.5422}{{\sf arXiv:1410.5422}}],
  [\href{http://dx.doi.org/10.1088/0264-9381/32/17/175008}{{\sf
  doi:10.1088/0264-9381/32/17/175008}}].

\bibitem{Cremmer:1980gs}
E.~Cremmer, {\it {Supergravities in 5 Dimensions}},  in {\em {In *Salam, A.
  (ed.), Sezgin, E. (ed.): Supergravities in diverse dimensions, vol. 1*
  422-437. (In *Cambridge 1980, Proceedings, Superspace and supergravity*
  267-282) and Paris Ec. Norm. Sup. - LPTENS 80-17 (80,rec.Sep.) 17 p. (see
  Book Index)}}, 1980.

\bibitem{Gunaydin:1983bi}
M.~Gunaydin, G.~Sierra, and P.~K. Townsend, {\it {The Geometry of N=2
  Maxwell-Einstein Supergravity and Jordan Algebras}},  {\sl Nucl. Phys.} {\bf
  B242} (1984) 244, [\href{http://dx.doi.org/10.1016/0550-3213(84)90142-1}{{\sf
  doi:10.1016/0550-3213(84)90142-1}}].

\bibitem{D'Auria:1981kq}
R.~D'Auria, E.~Maina, T.~Regge, and P.~Fre, {\it {Geometrical First Order
  Supergravity in Five Space-time Dimensions}},  {\sl Annals Phys.} {\bf 135}
  (1981) 237--269, [\href{http://dx.doi.org/10.1016/0003-4916(81)90155-X}{{\sf
  doi:10.1016/0003-4916(81)90155-X}}].

\bibitem{Khoury:2011da}
J.~Khoury, J.-L. Lehners, and B.~A. Ovrut, {\it {Supersymmetric Galileons}},
  {\sl Phys. Rev.} {\bf D84} (2011) 043521,
  [\href{http://arxiv.org/abs/1103.0003}{{\sf arXiv:1103.0003}}],
  [\href{http://dx.doi.org/10.1103/PhysRevD.84.043521}{{\sf
  doi:10.1103/PhysRevD.84.043521}}].

\bibitem{Farakos:2013zya}
F.~Farakos, C.~Germani, and A.~Kehagias, {\it {On ghost-free supersymmetric
  galileons}},  {\sl JHEP} {\bf 11} (2013) 045,
  [\href{http://arxiv.org/abs/1306.2961}{{\sf arXiv:1306.2961}}],
  [\href{http://dx.doi.org/10.1007/JHEP11(2013)045}{{\sf
  doi:10.1007/JHEP11(2013)045}}].

\bibitem{Srednicki:2007qs}
M.~Srednicki, {\em {Quantum field theory}}.
\newblock Cambridge University Press, 2007.

\bibitem{Freedman:2012zz}
D.~Z. Freedman and A.~Van~Proeyen, {\em {Supergravity}}.
\newblock Cambridge Univ. Press, Cambridge, UK, 2012.

\end{thebibliography}\endgroup

\end{document}